\documentclass[twoside,11pt]{article}

\usepackage[preprint]{jmlr2e}
\usepackage{caption}
\usepackage{subcaption}
\usepackage{amsmath}
\usepackage{xcolor}

\firstpageno{1}

\begin{document}

\title{Conditional Distribution Function Estimation Using Neural Networks for Censored and Uncensored Data}

\author{\name Bingqing Hu \email bingqih2@uci.edu \\
       \addr Department of Statistics\\
       University of California\\
       Irvine, CA 92697, USA
       \AND
       \name Bin Nan \email nanb@uci.edu \\
       \addr Department of Statistics\\
       University of California\\
       Irvine, CA 92697, USA}
       
\maketitle

\begin{abstract}
    Most work in neural networks focuses on estimating the conditional mean of a continuous response variable given a set of covariates.

    In this article, we consider estimating the conditional distribution function using neural networks for both censored and uncensored data. The algorithm is built upon the data structure particularly constructed for the Cox regression with time-dependent covariates. Without imposing any model assumption, we consider a loss function that is based on the full likelihood where the conditional hazard function is the only unknown nonparametric parameter, for which unconstraint optimization methods can be applied.       

    Through simulation studies, we show the proposed method possesses desirable performance, whereas the partial likelihood method and the traditional neural networks with $L_2$ loss yield biased estimates when model assumptions are violated. We further illustrate the proposed method with several real-world data sets. The implementation of the proposed methods is made available at \url{https://github.com/bingqing0729/NNCDE}.
\end{abstract}


\section{Introduction}

Neural networks are widely used in prediction. Depending on the nature of the outcome variable of interest, most of the current work falls into two categories: (i) estimating the conditional mean of a continuous response variable given a set of predictors, also called covariates, then predicting the outcome value given a new set of covariate values using the estimated conditional mean; (ii) estimating conditional probabilities of a categorical response variable given a set of covariates, then classifying a new data point only with known covariate values into one of the categories of the response variable. It is clearly seen that, from a statistical point of view, solving a classification problem is to estimate the conditional probability mass function, or equivalently, the conditional distribution function; whereas predicting a continuous outcome value often focuses on estimating a center of the conditional distribution function. In fact, estimating the conditional distribution function of a continuous response variable nonparametrically is of greater interest because it not only can determine the center of a distribution (mean or median), but also can set predictive intervals that are of great importance in prediction \citep{hall-jasa}. It is a challenge, however, to estimate the conditional distribution function given multiple covariates uisng traditional statistical methods \citep{hall-aos}. In this article, we propose to estimate the conditional distribution function of a continuous random variable given multiple covariates using neural networks .       

Our work is inspired by estimating the conditional survival function for censored failure time data, especially when covariates are time-dependent. In the survival analysis literature, a conditional survival function is usually estimated by fitting the semiparametric Cox proportional hazards model \citep{cox}. Although it has been widely used, particularly in health studies, the Cox model still requires strong model assumptions that can be violated in practical settings. Researchers have been exploring the application of neural networks  in survival analysis since 1990s. A line of research in the current literature uses neural networks to replace the linear component in a Cox model and the negative (logarithm of) partial likelihood as the loss function, see e.g.  \cite{faraggi} and \cite{XIANG2000243}.

Recently, \cite{deepsurv} revisited the Cox model and applied modern deep learning techniques (standardizing the input, using new activation function, new optimizer and learning rate scheduling) to optimizing the training of the network. Their neural networks model, called DeepSurv, outperforms Cox regressions evaluated by the C-index \citep{c_index} in several real data sets. In simulation studies, DeepSurv outperforms the Cox regression when the proportional hazards assumption is violated and performs similarly to the Cox regression when the proportional hazards assumption is satisfied. 

Following a similar modeling strategy, \cite{coxnnet} developed the neural networks model Cox-nnet for high-dimensional genetics data. Building upon the methodology of nested case-control studies, \cite{kvamme2019timetoevent} proposed to use a loss function that modifies the partial likelihood by subsampling the risk sets. Their neural networks models Cox-MLP(CC) and Cox-Time are computationally efficient and scale well to large data sets. Both models are relative risk models with the same partial likelihood but Cox-Time further removes the proportionality constraint by including time as a covariate into the relative hazard. Cox-MLP(CC) and Cox-Time were compared to the classical Cox regression, DeepSurv and a few other models on 5 real-world data sets and found to be highly competitive. We will compare our method to Cox-MLP(CC) and Cox-Time on 4 of the data sets where numbers of tied survival times are negligible. In all aforementioned methods, the semi-parametric nature of the model is retained, hence the baseline hazard function needs to be estimated in order to estimate the conditional survival function, whereas our method estimates the conditional hazard function directly without specifying any baseline hazard function. 

Another commonly used approach is to partition the time axis into a set of fixed intervals so that the survival probabilities are estimated on the set of discrete time points. \cite{biganzoli} proposed a model called PLANN in which the neural networks  contain a input vector of covariates and a discrete time point, and an output of the hazard rate 

at this time point. They used the negative logarithm of a Bernoulli likelihood as the loss function. \cite{street} proposed a model that outputs a vector with each element representing the survival probability at a predefined time point, and used a modified relative entropy error \citep{Solla1988AcceleratedLI} as the loss function. 

Similar to \cite{street}, \cite{brown} proposed a model that estimates hazard components which is defined as a value in $[0,1]$, the loss function is based on the sum of square errors.  More recent work includes Nnet-survival \citep{nnet} and RNN-SURV \citep{rnnsurv}. Both models require fixed and evenly partitioned time intervals, where Nnet-survival includes a convolutional neural network structure (CNN) and RNN-SURV uses a recurrent neural network structure (RNN). They all use Bernoulli type loss functions with differently created binary variables, where RNN-SURV adds an upper bound of the negative C-index into the loss function.

There are several limitations in the current literature of survival analysis using neural networks. Cox-type neural network models are still relative risk models with baseline hazards, which retain certain model structure, and the networks only output the relative risks. Moreover, these methods only deal with time-independent covariates. Methods using time partition have potential to incorporate time-varying covariates, but usually require fixed time partition.

Furthermore, in these methods, loss functions are often constructed heuristically.

To overcome these limitations, we propose a new method to estimate the conditional hazard function directly using neural networks. In particular, inspired by the standard data-expansion approach for the Cox regression with time-varying covariates,  we input time-varying covariates together with observed time points into a simple feed-forward network and output the logarithm of instantaneous hazard. We build the loss function from the logarithm of the full likelihood function, in which all functions, including the conditional hazard function, the conditional cumulative hazard function, and covariate processes, are evaluated only at the observed time points. Compared to existing methods, our new method has a number of advantages. First, we can handle time-varying covariates. Second, we make the least number of assumptions that only include conditional independent censoring  and that the instantaneous hazard given entire covariate paths only depends on values of covariates observed at the current time.
Third, estimating the (logarithm of) hazard function without imposing any constraint to the optimization automatically leads to a valid survival function estimator that is always monotone decreasing and bounded between 0 and 1.

Furthermore, since our loss function does not need to identify the risk set, scalable methods (e.g. training with batches in stochastic gradient descent) can be easily implemented to avoid blowing up the computer memory.

Estimating the conditional hazard function for censored survival data yields an estimation of the conditional survival function, hence equivalently the conditional distribution function, on the support of censored survival time given covariates. When there is no censoring, the problem naturally reduces to a general regression analysis, where the conditional distribution function is of interest. This clearly expands the scope of the current literature that primarily focuses on certain characteristic of the conditional distribution, for example, the conditional mean that corresponds to the mean regression. 

Once we obtain an estimator of the conditional distribution function, we can easily calculate the conditional mean 
given covariates, which provides a robust alternative approach to the mean regression using the $L_2$ loss. Note that the mean regression requires a basic assumption that the error term is uncorrelated with any of the covariates, which can be easily violated if some important covariate is unknown or unmeasured and correlated with some measured covariate. Our likelihood estimating approach, however, does not need such an assumption.

This article is organized as follows. We introduce our new methods in Section~\ref{sec:methods},    discuss simulation studies in Section~\ref{sec:simulation}, and illustrate comparisons to competing methods by analyzing several real world data sets in Section~\ref{sec:real data}. Finally, we provide   a few final remarks in Section~\ref{sec:discussion}.

\section{The Estimating Method Using Neural Networks}
\label{sec:methods}

In this section, we start with the likelihood-based loss function for estimating the conditional survival function given a set of time-varying covariates, then generalize the approach to estimating the conditional distribution function for an arbitrary continuous response variable, and finally provide an estimating procedure using neural networks. 

\subsection{Survival Analysis with Time-Varying Covariates}
\label{sec:survival}

\subsubsection{Data and Notation}

For subject $i$, we denote the time-varying covariate vector as $X_i(t)$, the underlying failure time as $T_i$, and the underlying censoring time as $C_i$, where $T_i$ possesses a Lebesgue density.  Let the observed time be $Y_i = \min\{T_i, C_i\}$ and the failure indicator be $\Delta_i = I(T_i\le C_i)$. We have $n$ independent and identically distributed (i.i.d.) observations $\{Y_i,\Delta_i,\widetilde X_i(\cdot): \ i=1,\dots,n\}$,  where $\widetilde X_i(t)$ denotes the covariate history up to time $t$, that is, $\widetilde X_i(t)=\{X_i(s), 0\le s \le t\}$. We assume each of the processes $X_i(t)$ has left continuous sample path with right limit. Let $f(t | \widetilde X_i(\infty))$ be the conditional Lebesgue density function of $T_i$, $f_C(t|\widetilde X_i(\infty))$ be the conditional density function of $C_i$, $S(t|\widetilde X_i(\infty))$ be the conditional survival function of $T_i$, and  $S_C(t|\widetilde X_i(\infty))$ be the conditional survival function of $C_i$.

Noting that the conditional survival probability given time-varying covaraites is not well-defined if there is an internal covariate, we assume that all covariates are external \citep{Kalbfleisch-Prentice-2002}. Specifically, the conditional hazard function of $T_i$ is independent of future covariate values and, furthermore, only depends on the current values of covariate processes:
$$
\lambda\left(t \left | \widetilde X_i(\infty)\right.\right)=\lambda\left(t \left | \widetilde X_i(t)\right.\right) = \lambda\left(t \left |X_i(t) \right.\right).
$$

Let $h(t, X_i(t)) = \log \lambda(t|X_i(t))$. Then the conditional cumulative hazard function given covariate history has the following form:
\begin{equation*}
    \label{eq:Lambda}
   \Lambda\left(t \left | \widetilde X_i(\infty)\right.\right)= \Lambda\left(t \left | \widetilde X_i(t)\right.\right)=\int_{0}^t \lambda\left(s \left | \widetilde X_i(t)\right.\right)ds=\int_{0}^t e^{h(s,X_i(s))}ds,
\end{equation*}
and the conditional survival function is given by
\begin{equation}
    \label{eq:S}
 S\left(t \left | \widetilde X_i(\infty) \right.\right)=   S\left(t \left | \widetilde X_i(t) \right.\right)=\exp\left\{-\Lambda \left(t \left | \widetilde X_i(t)\right)\right.\right\} = \exp\left\{-\int_{0}^t e^{h(s,X_i(s))}ds \right\}.
\end{equation}
Using $h$ instead of $\lambda$ in the above expression removes the positivity constraint for $\lambda$, hence simplifies the optimization algorithm for estimating the conditional survival function (\ref{eq:S}). Furthermore, equation (\ref{eq:S}) is always a valid survival function for any function $h$.

\subsubsection{Likelihood}

Assume censoring time is independent of failure time given covariates. Then given observed data   $\{y_i,\delta_i,x_i(\cdot)\}$, $i=1,\dots,n$, the full likelihood function becomes

\begin{eqnarray*}
    \label{eq:likelihood}
        L_n &=& \prod_{i=1}^n \left\{f(y_i | \widetilde x_i(y_i))S_C(y_i | \widetilde x_i(y_i))\right\}^{\delta_i}\left\{f_C(y_i | \widetilde x_i(y_i))S(y_i | \widetilde x_i(y_i))\right\}^{1-\delta_i} \nonumber\\
   %     &=&\prod_{i=1}^n \lambda(y_i \mid x_i(y_i))^{\delta_i}S_i(y_i \mid x_i(y_i)) f_C(y_i \mid x_i(y_i))^{1-\delta_i}S_C(y_i \mid x_i(y_i))^{\delta_i} \nonumber \\
        &\propto& \prod_{i=1}^n \lambda(y_i | x_i(y_i))^{\delta_i}S(y_i |\widetilde x_i(y_i)) \nonumber \\ 
        &=& \prod_{i=1}^n \exp\{h(y_i,x_i(y_i))\delta_i\}\exp\left\{-\int_0^{y_i} e^{h(t,x_i(t))}dt \right\}.
\end{eqnarray*}

Thus, the log likelihood is given by 
\begin{equation} \label{eq:log-likelihood}
        \ell_n = \sum_{i=1}^n \left\{h(y_i,x_i(y_i))\delta_i-\int_0^{y_i} e^{h(t,x_i(t))}dt\right\}.
\end{equation}

\subsubsection{Data Structure and Discretized Loss}

\begin{table}[p] %!tbh
    \centering
    \begin{tabular}{ccccc}
    \hline
    $i$ & start time & stop time & $\delta_{ij}$ & covariates\\
    \hline
    1 & $t_0=0$ & $t_1$ & 0 & $x_1(t_1)$\\
    1 & $t_1$ & $t_2$ & 0 & $x_1(t_2)$\\
    \vdots & \vdots & \vdots & \vdots & \vdots\\
    1 & $\cdots$ & $y_1$ & $\delta_1$ & $x_1(y_1)$ \\
    2 & $t_0=0$ & $t_1$ & 0 & $x_2(t_1)$\\
    2 & $t_1$ & $t_2$ & 0 & $x_2(t_2)$\\
    \vdots & \vdots & \vdots & \vdots & \vdots \\
    2 & $\cdots$ & $y_2$ & $\delta_2$ & $x_2(y_2)$\\
    \vdots &  \vdots & \vdots & \vdots & \vdots\\
    \hline
    \end{tabular}
    \caption{The expanded data set for survival problem with time varying covariate, where  $(t_1, \dots, t_n)$ are sorted values of $(y_1, \dots, y_n)$ from the training set.}
    \label{tab:data_structure_surv}
\end{table}

When fitting the Cox model with time-varying covariates, the data set is usually expanded to the structure given in Table~\ref{tab:data_structure_surv} so each row is treated as an individual observation, where $(t_1, \dots, t_n)$ are sorted values of observed times $(y_1, \dots, y_n)$ and $\delta_{ij} = \delta_iI(t_j=y_i)$. Specifically, the time axis is partitioned naturally by observed times. The same data structure can be applied to maximizing the log likelihood function (\ref{eq:log-likelihood}) at the grid points $(t_1, \dots, t_n)$, or equivalently, minimizing the following loss function:

\begin{equation}
%\label{eq:loss_disc}
\label{eq:loss}
        loss(h) = \frac{1}{n} \sum_{i=1}^n \sum_{j=1}^n I(t_j \le y_i) \left\{ e^{h(t_j,x_i(t_j))}(t_j - t_{j-1}) -h(t_j,x_i(t_j))\delta_{ij}\right\},
\end{equation}
where $t_0=0$. It becomes clear that the expanded data set in Table \ref{tab:data_structure_surv} provides a natural way of implementing numerical integration in the negative log likelihood $-n^{-1}\ell_n$  based on empirical data. Once an estimator of $h$ is obtained using neural networks (see Subsection \ref{sec:hyper} for details), the conditional survival function can be estimated by plugging the estimated $h$ into equation (\ref{eq:S}).

\subsection{Estimation of Conditional Distribution for Uncensored Data}

If there is no censoring, then $\delta_i=1$ for all $i \in \{1, \dots, n\}$ in the log likelihood function (\ref{eq:log-likelihood}).  Now consider an arbitrary continuous response variable $Y \in (-\infty, \infty)$ that is no longer ``time." Note that the time variable $T \in [0, \infty)$. We are interested in estimating $F(y|x)$, the conditional distribution function of $Y$ given covariates $X=x$, where $X$ is a random vector. Since there is no time component in general, covariates are no longer ``time-varying." 

Assume $\{Y_i, X_i\}$, $i=1, \dots, n$, are i.i.d. Denote the observed data as $\{y_i, x_i\}$, $i=1, \dots, n$. 

We generalize the idea of using hazard function in survival analysis to estimate $F(y|x)$ for an arbitrary continuous random variable $Y$. Again, let  $\lambda(t | x_i)=e^{h(t,x_i)}$. Then the conditional cumulative hazard function becomes
\begin{equation*}
    \label{eq:general_Lambda}
    \Lambda(t |  x_i)=\int_{-\infty}^t \lambda(s |  x_i)ds=\int_{-\infty}^t e^{h(s,x_i)}ds,
\end{equation*} 
which gives the conditional distribution function
\begin{equation}
    \label{eq:CDF}
    F(t |  x_i)=1-\exp\left\{-\Lambda(t | x_i)\right\}.
\end{equation}
Hence, the log likelihood function is given by
\begin{equation} \label{eq:general_loglikelihood}
        \ell_n = \sum_{i=1}^n \left\{h(y_i,x_i)-\int_{-\infty}^{y_i} e^{h(t,x_i)}dt\right\}.
\end{equation}

Note that the above log likelihood has the same form as (\ref{eq:log-likelihood}) except that the covariates are not time-varying, $\delta_i = 1$ for all $i$, and integrals start from $-\infty$. As a way of evaluating integrals in the log likelihood, the expanded data structure in Table \ref{tab:data_structure_surv} can be useful in estimating $h(y,x)$ with slight modifications given in Table \ref{tab:data_structure_general}.

\begin{table}[!htp]
    \centering
    \begin{tabular}{ccccc}
    \hline
    $i$ & start & stop & $\delta_{ij}$ & covariates\\
    \hline
    1 & $t_0=-\infty$ & $t_1$ & 0 & $x_1$\\
    1 & $t_1$ & $t_2$ & 0 & $x_1$\\
    \vdots & \vdots & \vdots & \vdots & \vdots\\
    1 & ... & $y_1$ & 1 & $x_1$  \\
    2 & $t_0=-\infty$ & $t_1$ & 0 & $x_2$\\
    2 & $t_1$ & $t_2$ & 0 & $x_2$\\
    \vdots & \vdots & \vdots & \vdots & \vdots \\
    2 & ... & $y_2$ & 1 & $x_2$\\
    \vdots &  \vdots & \vdots & \vdots & \vdots\\
    \hline
    \end{tabular}
    \caption{The expanded data set for estimating the conditional distribution function of a continuous response variable, where $(t_1,\dots,t_n)$ are sorted values of $(y_1,\dots,y_n)$ from the training set.}
    \label{tab:data_structure_general}
\end{table}

To be numerically tractable, we assign $1/n$ to be the value of the distribution function at $t_1$, in other words, we make $F(t_1 | x_i)=1/n$, which is the empirical probability measure of $Y$ at $t_1$. Thus $\Lambda(t_1 | x_i) = -\log(1-1/n)$. Letting $\delta_{ij}=I(t_j=y_i)$ and evaluating the integrals in (\ref{eq:general_loglikelihood}) on grid points $(t_1, \dots, t_n)$ that are sorted values of $(y_1, \dots, y_n)$, we obtain the following loss function:

\begin{eqnarray}
        loss(h) &=& \frac{1}{n} \sum_{i=1}^n \left\{ -\log(1-1/n)+\sum_{j=2}^n I(t_j \le y_i) \left[ e^{h(t_j,x_i)}(t_j-t_{j-1}) - h(t_j,x_i)\delta_{ij} \right] \right\} \nonumber \\
        &=& \frac{1}{n} \sum_{i=1}^n \sum_{j=2}^n I(t_j \le y_i)\left\{ e^{h(t_j,x_i)}(t_j-t_{j-1}) - h(t_j,x_i)\delta_{ij}\right\} + \rm{Constant}. \label{eq:loss-general}
\end{eqnarray}
Once an estimator of $h$, denoted by $\widehat h$, is obtained, the conditional distribution function (\ref{eq:CDF}) can be estimated by 
\begin{equation} \label{eq:CDF-Estimator}
\widehat F(y|x) = I(t_1 \le y)\left[1- \frac{n-1}{n}\exp\left\{ - \sum_{j=2}^n I(t_j \le y) e^{\widehat h(t_j, x)} (t_j - t_{j-1}) \right\} \right].
\end{equation}

\begin{remark}
If the support of the continuous response variable has a fixed finite lower bound, then the integration for the conditional cumulative hazard function is the same as that for survival data. In other words, there is no need to assign a point mass of $1/n$ at $t_1$. 
\end{remark}

\subsection{Neural Networks, Hyperparameters and Regularization} 
\label{sec:hyper}

We propose to estimate the arbitrary function $h(t,x_i(t))$, or $h(t,x_i)$ when covariates are not time-varying, by minimizing the respective loss function (\ref{eq:loss}) or (\ref{eq:loss-general}) using neural networks. We then obtain the estimated conditional survival curve or the conditional distribution function from Equation (\ref{eq:S}) or Equation (\ref{eq:CDF-Estimator}), respectively. The input of neural networks  is $(t_{j-1}, t_j, x_i(t_j))$ or $(t_{j-1}, t_j, x_i)$ in each row of Table \ref{tab:data_structure_surv} or Table \ref{tab:data_structure_general}, and the output is $\widehat h(t,x_i(t))$ or $\widehat h(t,x_i)$,  respectively. Note that the first row for each $i$ in Table \ref{tab:data_structure_general} is excluded from the calculation.

\begin{figure}[p]
    \centering
    \includegraphics[width=0.8\textwidth]{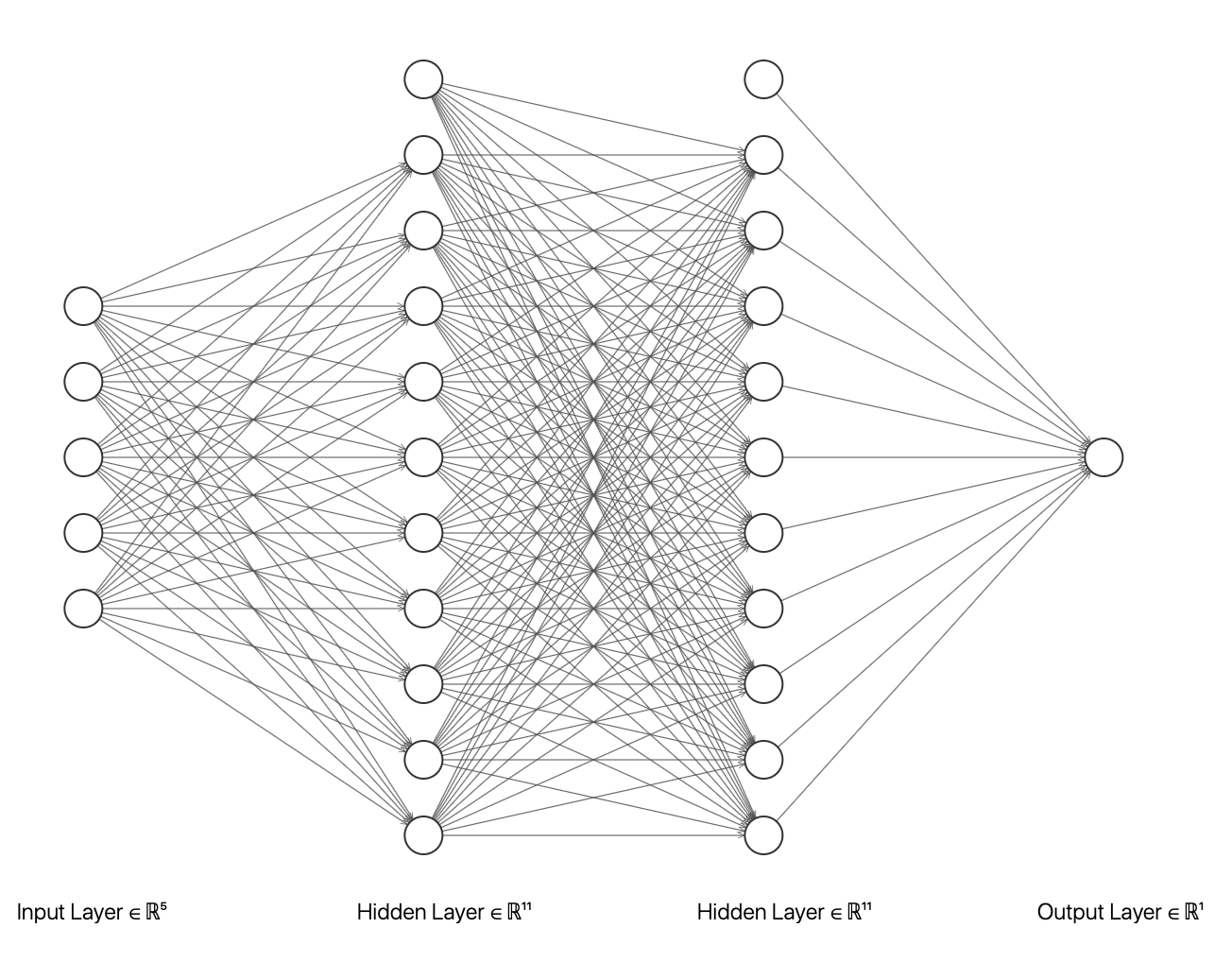}
    \caption{An example of fully connected feed forward neural networks with 2 hidden layers. In this example, the input dimension is 4 plus an intercept term, each hidden layer contains 10 nodes plus an intercept node and the output is a single value.}
    \label{fig:fcnn}
\end{figure}

We use tensorflow.keras \citep{chollet2015keras} to build and train the neural networks. The network structure is a fully connected feed forward neural network  with two hidden layers and a single output value. The input layer consists of $t_{j-1}$, $t_j$, and covariates. In addition, an intercept term is included in each layer (see Figure~\ref{fig:fcnn}). The relu function is used as the activation function between input and hidden layers, and the linear function is used for the output so that the output value is not constrained. We use Adam \citep{adam} as the optimizer. Other hyperparameters include the number of nodes in each layer, the batch size and the initial learning rate. In our simulations, the number of nodes in each hidden layer is 64, the batch size is 100, and the initial learning rate is 0.001. To have a fair comparison in real-world data examples, we tune the hyperparameters from the set of all combined values with the number of nodes in each hidden layer taken from $\{64, 128, 256\}$, the initial learning rate from $\{0.1, 0.01, 0.001, 0.0001\}$, and the batch size from $\{64, 128, 256\}$.

We use early stopping to avoid over-fitting. According to \cite{Goodfellow-et-al-2016}, early stopping has the advantage over explicit regularization methods in that it automatically determines the correct amount of regularization. 

Specifically, we randomly split the original data into training set and validation set with 1:1 proportion. When the validation loss is no longer decreasing in 10 consecutive steps, we stop the training. To fully use the data, we fit the neural networks  again by swapping the training and the validation sets, then average both outputs as the final result.

\section{Simulations}
\label{sec:simulation}

\subsection{Censored Data with Time-Varying Covariates}
\label{sec:sim survival}

For censored survival data with time-varying covariates, we aim to compare our method of using neural networks  to the partial likelihood method for the Cox model under two different setups. 

In the first setup, we generate data following the proportional hazards assumption so the Cox model is the gold standard. In the second setup, we generate data from the model with a quadratic term and an interaction term in the log relative hazard function so the Cox model with original covariates in the linear predictor becomes a misspecified model. Details are given below.

\begin{enumerate}
    \item In both setups, we use the hazard function of a scaled beta distribution as the baseline hazard: 
    \begin{equation*}
        \lambda_0(t)=\frac{f_0(t/\tau)}{1-F_0(t/\tau)},
    \end{equation*}
    where $f_0(.)$ and $F_0(.)$ are the density and the distribution functions of $\rm{Beta} \, (8,1)$. We use $\tau=100$ so that $t \in [0,100]$. 
    \item Generate time-varying covariates on a fine grid of $[0,\tau]$. 
    For $t \in \{0, \Delta s, 2\Delta s,....,\tau\}$ with $\Delta s = 0.01$, $i \in \{1,2,...n\}$, we generate random variables $\alpha_{i1},\dots,\alpha_{i5}$ independently from a $\rm{Uniform} \, (0,1)$ distribution, and $q_i$ independently from a $\rm{Uniform} \, (0,\tau)$ distribution, and construct two time-varying  covariates as follows:
\begin{eqnarray*}    
    x_{1i}(t)&=&\alpha_{i1}+\alpha_{i2}\sin(2\pi t/\tau)+\alpha_{i3}\cos(2\pi t/\tau)+\alpha_{i4}\sin(4\pi t/\tau)+\alpha_{i5}\cos(4\pi t/\tau), \\
    x_{2i}(t)&=& 
     \begin{cases}
  0, & \mbox{if } t \le q_i; \\
  1, & \mbox{if } t > q_i.
  \end{cases}
 \end{eqnarray*}
The sample paths of both covariates are left-continuous step functions with right limit.  We also generate three time-independent covariates:
\begin{eqnarray*}
  x_{3i} &\sim& \rm{Bernoulli} \, (0.6), \\
  x_{4i} &\sim& \rm{Poisson} \, (2), \mbox{truncated at } 5, \\
  x_{5i} &\sim& \rm{Beta} \, (2,5). 
\end{eqnarray*}
    \item In {\em Setup 1}, the conditional hazard function is
    \begin{equation} \label{eq: setup1}
        \lambda(t | x_i(t))=\lambda_0(t)e^{2x_{1i}(t)+2x_{2i}(t)+2x_{3i}+2x_{4i}+2x_{5i}}, 
    \end{equation}
and in {\em Setup 2}, 
    \begin{equation} \label{eq:setup2}
        \lambda(t | x_i(t))=\lambda_0(t)e^{2x_{1i}(t)^2+2x_{2i}(t)+2x_{3i}x_{4i}+2x_{5i}}.
    \end{equation}
Clearly, fitting the Cox model $\lambda(t|x_i(t)) = \lambda_0(t) \exp\{\beta_1 x_{1i}(t) + \beta_2 x_{2i}(t) + \beta_3 x_{3i} + \beta_4 x_{4i} + \beta_5 x_{5i} \}$ with data generated from (\ref{eq:setup2}) in Setup 2 will not yield desirable results. 

    \item %Since the mathematical evaluation is complicated, 
    Once covariates are generated, we numerically evaluate the conditional cumulative hazard function and the conditional survival function on the fine grid of survival time. Specifically, for $s \in \{0, \Delta_s, 2\Delta s, \dots, \tau\}$,  
    \begin{eqnarray*}
        \Lambda(t |\widetilde x_i(t)) &=& \Delta s \sum_{s\le t}\lambda_i(s | x_i(s)), \\
        S(t |\widetilde x_i(t)) &=& \exp\left\{-\Lambda_i(t | \widetilde x_i(t))\right\}.
    \end{eqnarray*}

    \item For $i \in \{1,2,...n\}$, we generate random variable $u_i$ from a $\rm{Uniform}\, (0,1)$ distribution, then obtain the failure time by $t_i=\sup\left\{t:S_i(t |\widetilde x_i(t)) \ge u_i\right\}$.
    \item We generate the censoring time $c_i$ from an $\rm{Exponential} \, (100)$ distribution. Then we have  $y_i = t_i \wedge c_i$ and $\delta_i=I(t_i\le c_i)$. In both setups (\ref{eq: setup1}) and (\ref{eq:setup2}),  censoring rates are around 20\%.
\end{enumerate}

For each simulation setup, we independently generate a training set and a validation set with equal sample size, then fit our model using neural networks . We refit the model by swapping training and validation sets, and take the average as our estimator. For the Cox regression, we maximize the partial likelihood using all data. We repeat the process for $N$ independent data sets, and calculate the average and sample variance of these $N$ estimates at each time point on the fine grid for another set of randomly generated covariates. Finally, we plot the sample average of conditional survival curves estimated by neural networks  together with the empirical confidence band, the average conditional survival curves estimated from the Cox regression, and the true conditional survival curve for a comparison.

%\subsubsection{Results}

\begin{figure}[p]
\centering
\begin{subfigure}{0.3\textwidth}
\includegraphics[width=0.9\linewidth, height=4cm]{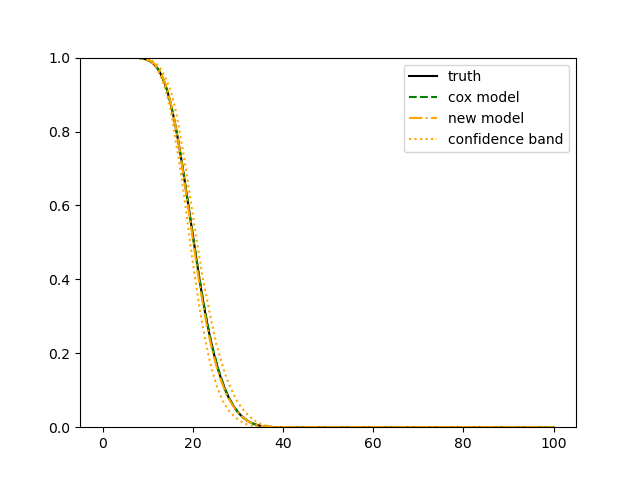} 
\end{subfigure}\hfil
\begin{subfigure}{0.3\textwidth}
\includegraphics[width=0.9\linewidth, height=4cm]{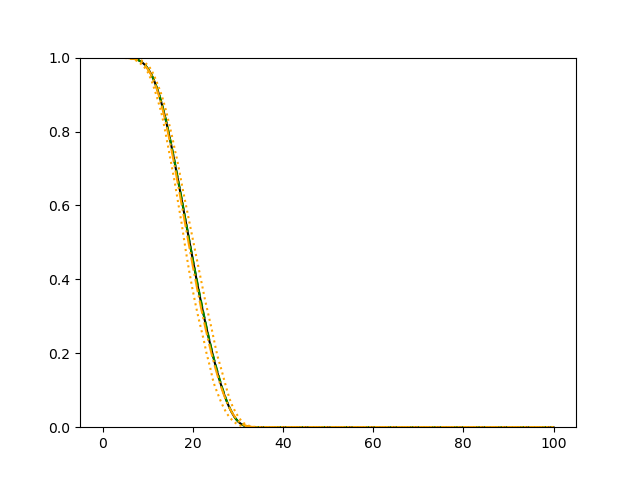}
\end{subfigure}\hfil
\begin{subfigure}{0.3\textwidth}
\includegraphics[width=0.9\linewidth, height=4cm]{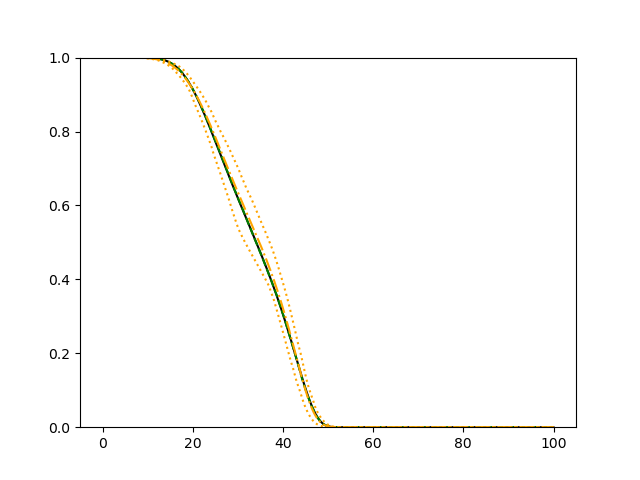}
\end{subfigure}\hfil
\begin{subfigure}{0.3\textwidth}
\includegraphics[width=0.9\linewidth, height=4cm]{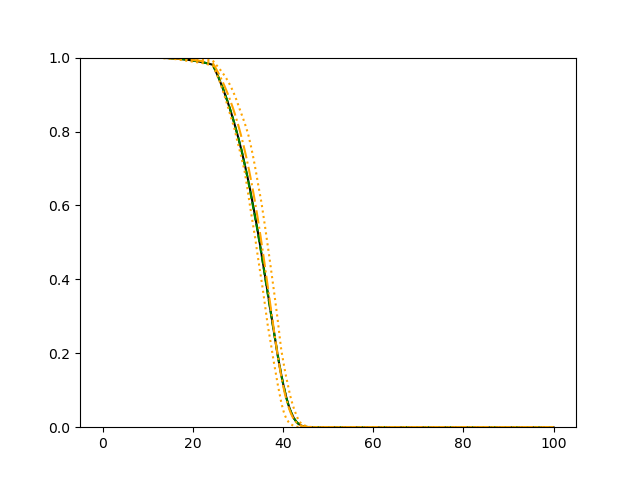}
\end{subfigure}\hfil
\begin{subfigure}{0.3\textwidth}
\includegraphics[width=0.9\linewidth, height=4cm]{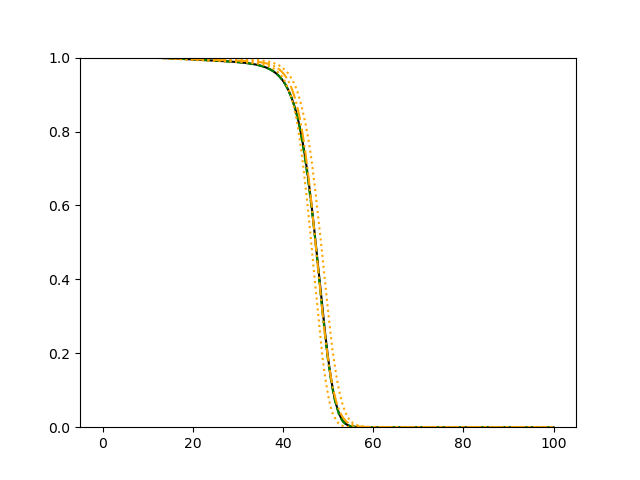} 
\end{subfigure}\hfil
\begin{subfigure}{0.3\textwidth}
\includegraphics[width=0.9\linewidth, height=4cm]{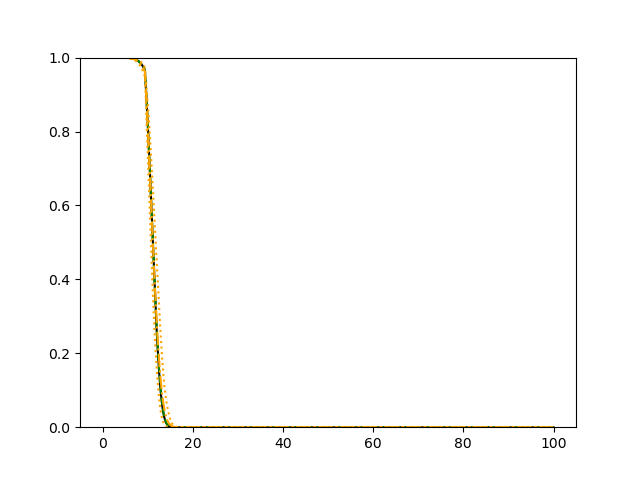} 
\end{subfigure}\hfil
\begin{subfigure}{0.3\textwidth}
\includegraphics[width=0.9\linewidth, height=4cm]{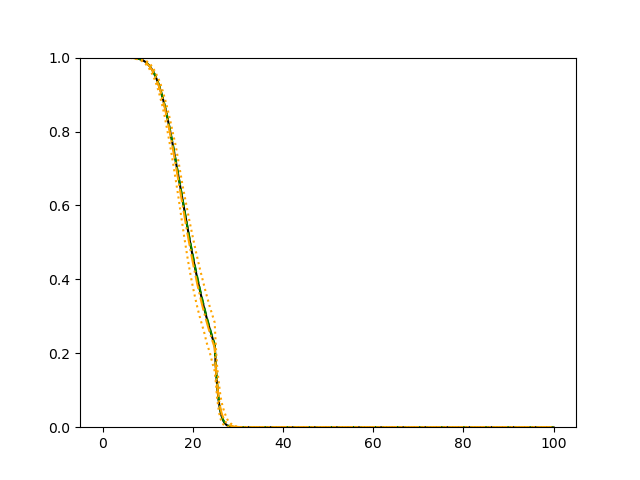} 
\end{subfigure}\hfil
\begin{subfigure}{0.3\textwidth}
\includegraphics[width=0.9\linewidth, height=4cm]{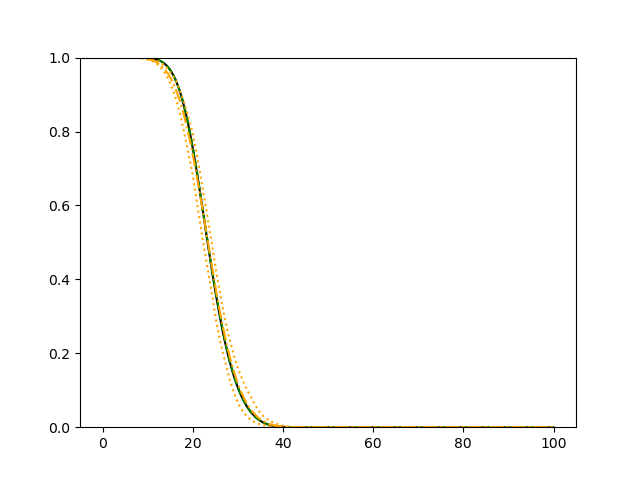} 
\end{subfigure}\hfil
\begin{subfigure}{0.3\textwidth}
\includegraphics[width=0.9\linewidth, height=4cm]{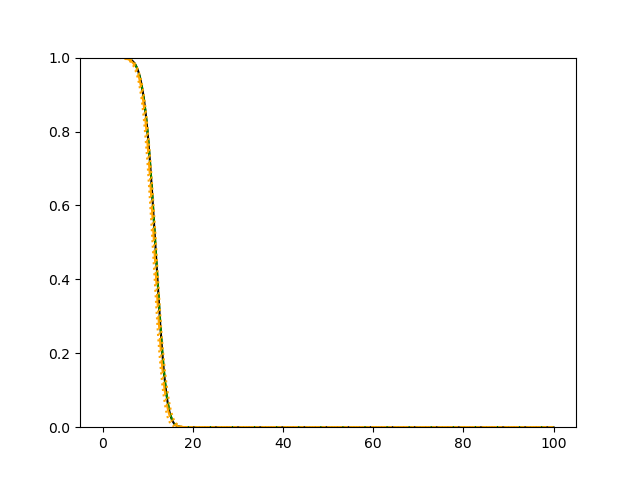} 
\end{subfigure}\hfil
\caption{Conditional survival curves for 9 different sets of covariates when the Cox model is corrected specified. }
\label{fig:cox}
\end{figure}

\begin{figure}[p] 
\centering
\begin{subfigure}{0.3\textwidth}
\includegraphics[width=0.9\linewidth, height=4cm]{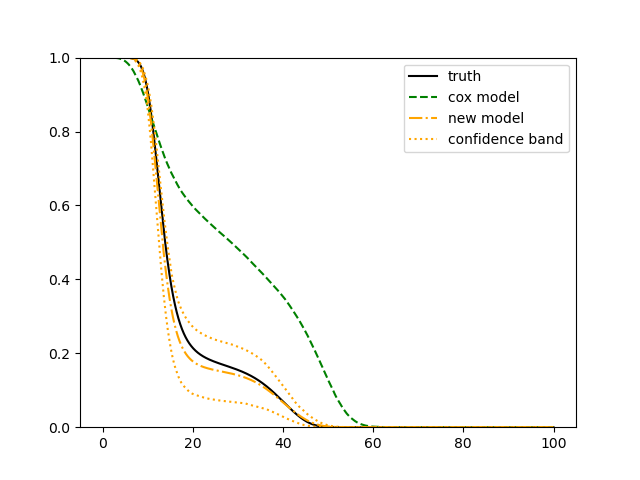} 
\end{subfigure}\hfil
\begin{subfigure}{0.3\textwidth}
\includegraphics[width=0.9\linewidth, height=4cm]{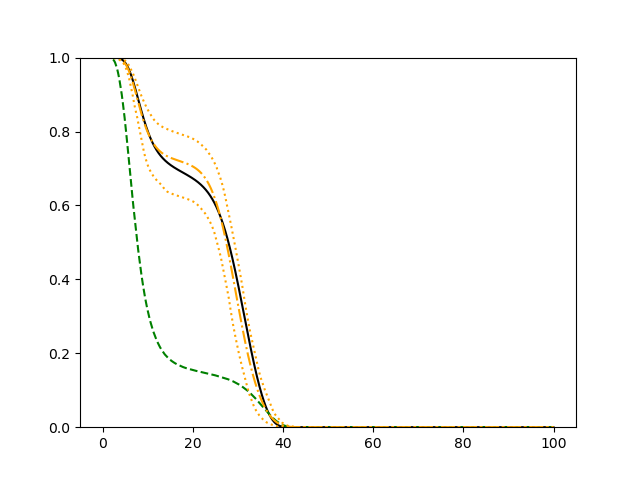}
\end{subfigure}\hfil
\begin{subfigure}{0.3\textwidth}
\includegraphics[width=0.9\linewidth, height=4cm]{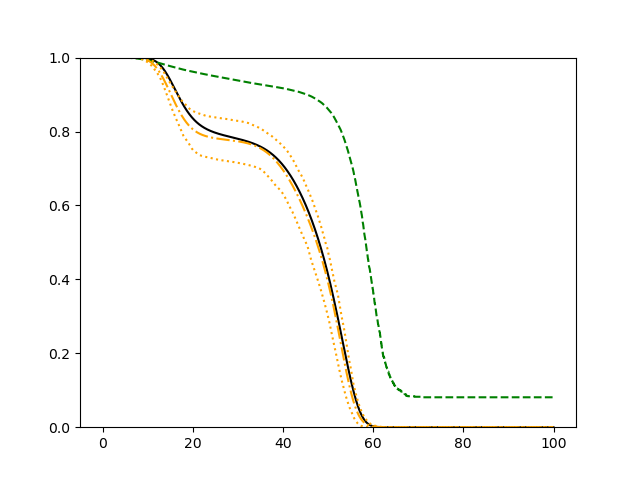}
\end{subfigure}\hfil
\begin{subfigure}{0.3\textwidth}
\includegraphics[width=0.9\linewidth, height=4cm]{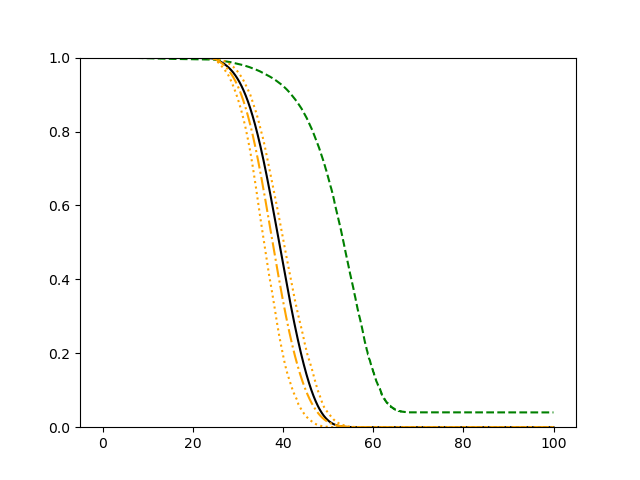}
\end{subfigure}\hfil
\begin{subfigure}{0.3\textwidth}
\includegraphics[width=0.9\linewidth, height=4cm]{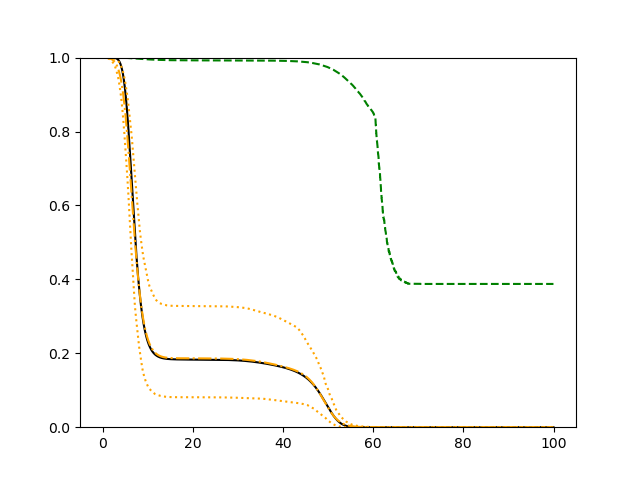} 
\end{subfigure}\hfil
\begin{subfigure}{0.3\textwidth}
\includegraphics[width=0.9\linewidth, height=4cm]{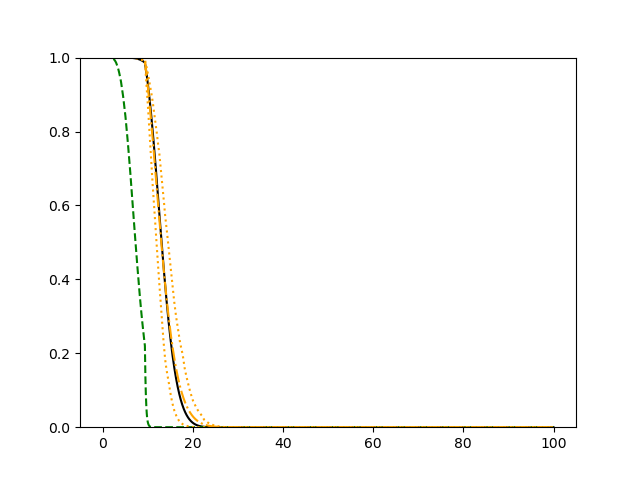} 
\end{subfigure}\hfil
\begin{subfigure}{0.3\textwidth}
\includegraphics[width=0.9\linewidth, height=4cm]{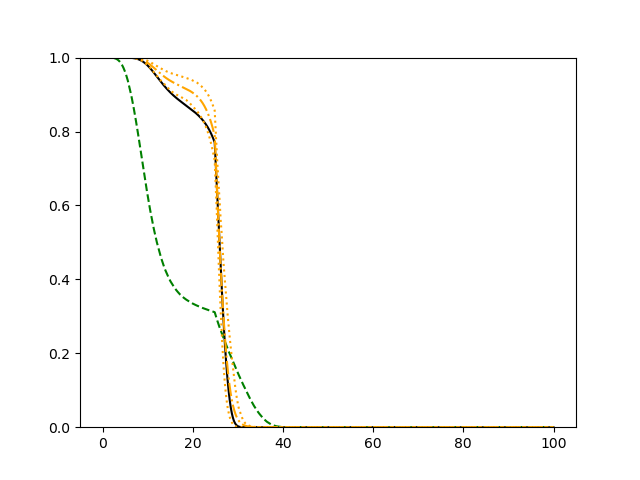} 
\end{subfigure}\hfil
\begin{subfigure}{0.3\textwidth}
\includegraphics[width=0.9\linewidth, height=4cm]{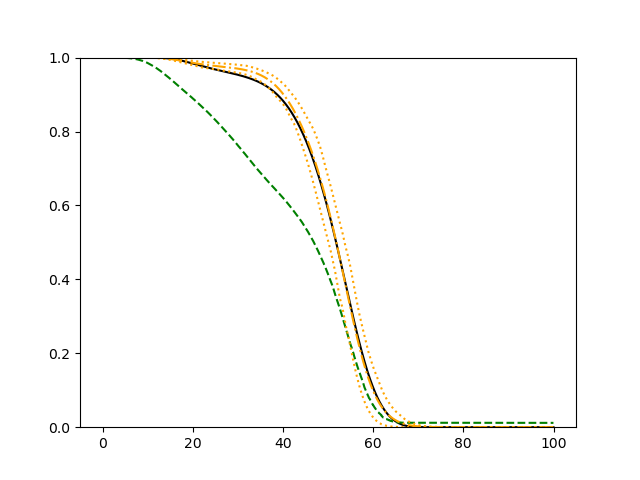} 
\end{subfigure}\hfil
\begin{subfigure}{0.3\textwidth}
\includegraphics[width=0.9\linewidth, height=4cm]{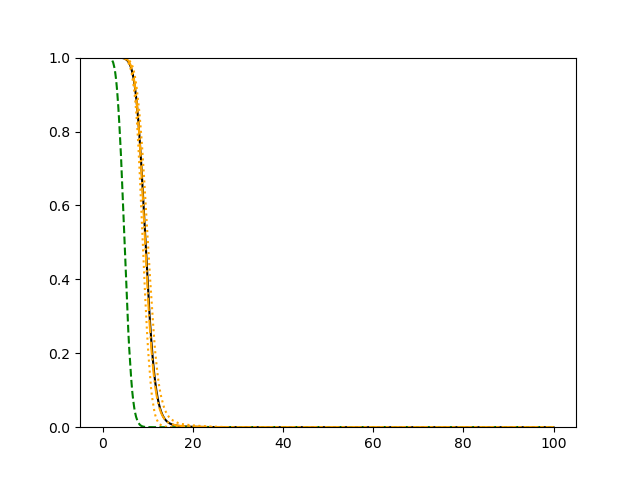} 
\end{subfigure}\hfil

\caption{Conditional survival curves for 9 different sets of covariates when the Cox proportional hazards assumption is violated.}
\label{fig:non-cox}
\end{figure}

The simulation results illustrated in both Figure~\ref{fig:cox} and Figure~\ref{fig:non-cox} are based on a sample size of $n=3000$ (1500 for training and 1500 for validation) with 100 repetitions, where the curves for 9 different sets of covariates are presented. The green dashed line is the average estimated curve by using the partial likelihood method,  the orange dash-dot line is the average estimated curve by our proposed neural networks  method, and  the black solid line is the truth curve.
The dotted orange curves are the 90\% confidence band obtained from the 100 repeated simulation runs using the proposed method.
From Figure~\ref{fig:cox} we see that when the Cox model is correctly specified, both the partial likelihood method and our proposed neural networks  method perform well, with estimated survival curves nicely overlapping with the corresponding true curves.  When the  Cox model is misspecified, Figure~\ref{fig:non-cox} shows that the partial likelihood approach yields severe biases, whereas the proposed neural networks  method still works well with a similar performance to that in Setup 1 shown in Figure~\ref{fig:cox} .

\subsection{Uncensored Data}
\label{sec:sim general}

For uncensored continuous outcomes, the traditional neural networks  method with the commonly used  $L_2$ loss function gives the conditional mean estimator. Then the conditional distribution function given a set of covariate values can be estimated by shifting the center of the empirical distribution of training set residuals to the estimated conditional mean. This would yield a valid estimator under the assumption that the errors (outcomes subtract their conditional means) are i.i.d. and uncorrelated with conditional means. We will evaluate the impact of this widely imposed condition for the mean regression via simulations. On the other hand, an estimator of the conditional distribution function gives a conditional mean estimator as follows:
\begin{equation*}
    \int_{-\infty}^{\infty}td\widehat F(t|x) = \sum_{i=1}^n t_i \left(\widehat F(t_i|x) - \widehat F_k(t_{i-1}|x)\right).
\end{equation*}
Thus, we will compare our method to the method with $L_2$ loss on the estimation of the conditional distribution function as well as the estimation of the conditional mean.

In the following simulation studies, we consider i.i.d. data generated from the following model:
$$y_i = x_{1i}^2+x_{2i}x_{3i}+x_{3i}x_{4i}+x_{5i}+\epsilon_i g(x_i),$$
$i=1, \dots, n$, where $x_i$ denotes the $i$-th vector of all covariates, $\epsilon_i$ is mean-zero given all covariates, and $g$ is a function of covariates, so $\epsilon_i g(x_i)$ is the $i$-th error term with zero-mean. We consider two simulation setups. In the first setup, the error is uncorrelated with the mean and has constant variance. In the second setup, the error is correlated with the mean and has non-constant variance. We would expect our new method outperforms the method with $L_2$ loss since our loss function is based on the nonparametric likelihood function that is free of any model assumption. Specifically, covariate values $x_{1i}, \dots, x_{5i}$ are generated from the following distributions:
\begin{eqnarray*}
x_{1i} &\sim& \rm{N} \, (0,1), \mbox{ truncated at } \pm 3, \\ 
    x_{2i} &\sim&  \rm{Uniform} \, (0,1), \\
    x_{3i} &\sim& \rm{Beta} \, (0.5,0.5),\\
    x_{4i} &\sim&  \rm{Bernoulli} \, (0.5), \\
    x_{5i} &\sim& \rm{Poisson} \, (2), \mbox{ truncated at } 5. \\
\end{eqnarray*}
And the two setups are:
\begin{itemize}
\item[]{\em Setup 1} (uncorrelated error with constant variance): generate another covariate $x_{6i} \sim \rm{N}\,(1,1)$ independently, then generate $\epsilon_i \sim$ a mixture distribution of $\rm{N} \, (-2,1)$, $\rm{N} \, (0,1)$,  and  $0.5 x_{6i}^2$ with mixture probabilities $(0.1, 0.7, 0.2)$, and let $g(x_i)=c$, where $c$ is a constant.

\item[] {\em Setup 2} (correlated error with non-constant variance): generate another covariate $x_{6i} \sim {\rm N}\,(1+0.5 x_{1i},0.75)$, such that 
\[\begin{pmatrix}
x_{1i} \\
x_{6i}
\end{pmatrix}\sim \rm{N} \left(\begin{pmatrix}
0 \\
1
\end{pmatrix},\begin{pmatrix}
1 & 0.5 \\
0.5 & 1
\end{pmatrix}\right),
\]
then generate $\epsilon_i \sim$ a mixture distribution of $\rm{N} \, (-2,1)$, $\rm{N} \, (0,1)$,  and  $0.5 x_{6i}^2$ with mixture probabilities $(0.1, 0.7, 0.2)$, and let 
    $g(x_i)=cx_{1i}^2$, where $c$ is a constant.

\end{itemize}
Note that different values of constant $c$ yield different signal to noise ratios in both setups.

\begin{figure}[p] 
\centering
\begin{subfigure}{0.3\textwidth}
\includegraphics[width=0.9\linewidth, height=4cm]{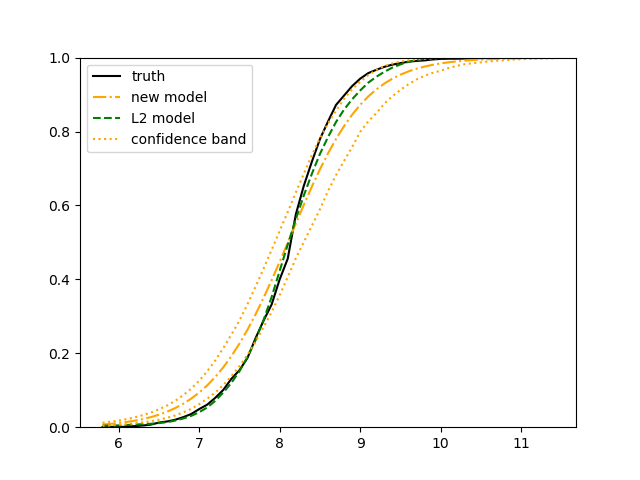} 
\end{subfigure}\hfil
\begin{subfigure}{0.3\textwidth}
\includegraphics[width=0.9\linewidth, height=4cm]{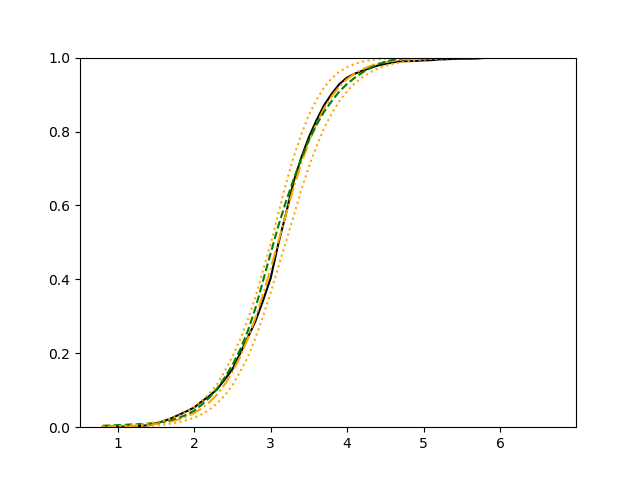}
\end{subfigure}\hfil
\begin{subfigure}{0.3\textwidth}
\includegraphics[width=0.9\linewidth, height=4cm]{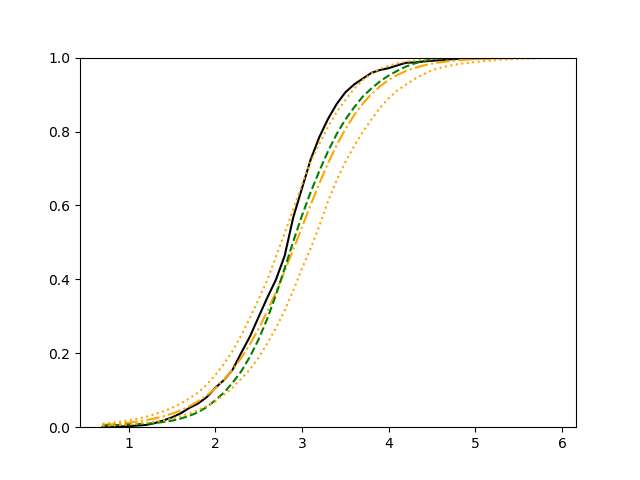}
\end{subfigure}\hfil
\begin{subfigure}{0.3\textwidth}
\includegraphics[width=0.9\linewidth, height=4cm]{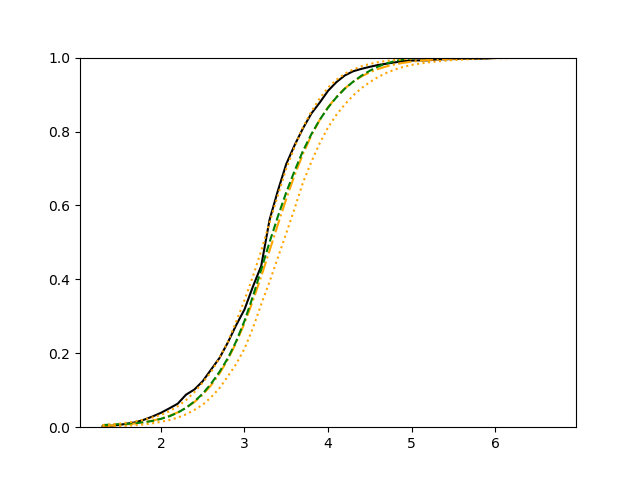}
\end{subfigure}\hfil
\begin{subfigure}{0.3\textwidth}
\includegraphics[width=0.9\linewidth, height=4cm]{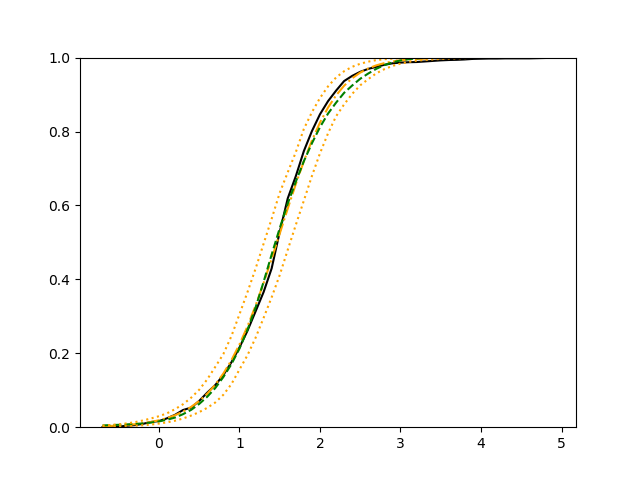} 
\end{subfigure}\hfil
\begin{subfigure}{0.3\textwidth}
\includegraphics[width=0.9\linewidth, height=4cm]{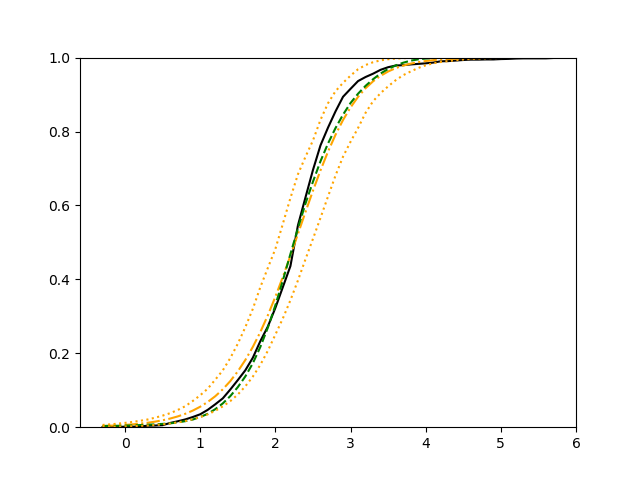} 
\end{subfigure}\hfil
\begin{subfigure}{0.3\textwidth}
\includegraphics[width=0.9\linewidth, height=4cm]{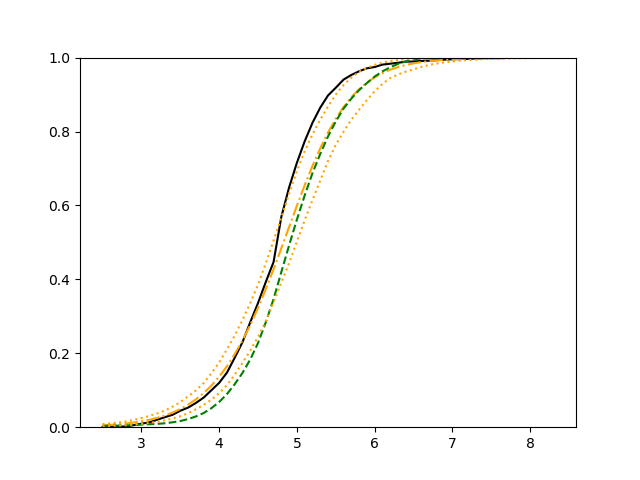} 
\end{subfigure}\hfil
\begin{subfigure}{0.3\textwidth}
\includegraphics[width=0.9\linewidth, height=4cm]{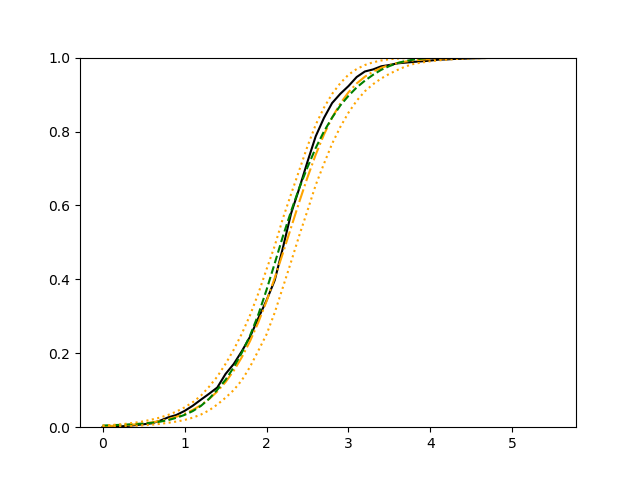} 
\end{subfigure}\hfil
\begin{subfigure}{0.3\textwidth}
\includegraphics[width=0.9\linewidth, height=4cm]{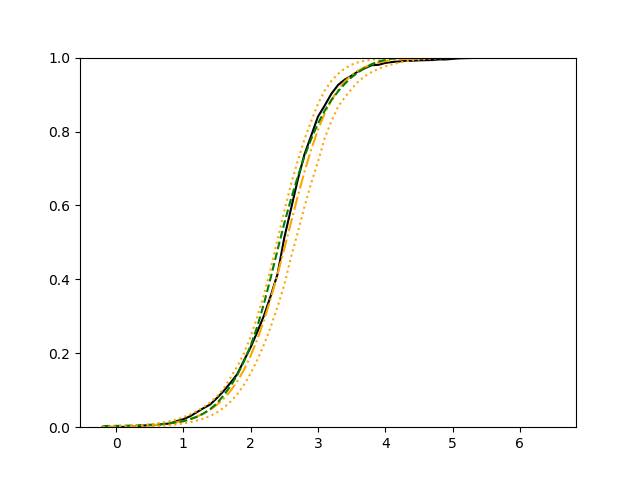} 
\end{subfigure}\hfil
\caption{Conditional distribution functions given 9 different sets of covariate values for uncensored data generated in Setup 1 with c=0.5.}
\label{fig:l2}
\end{figure}

\begin{figure}[p]
\centering
\begin{subfigure}{0.3\textwidth}
\includegraphics[width=0.9\linewidth, height=4cm]{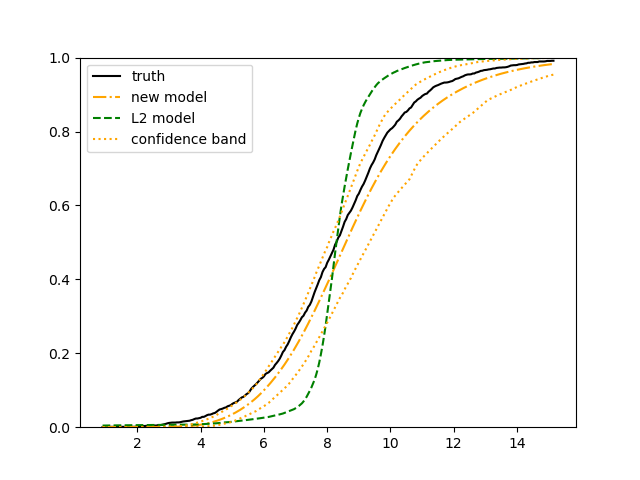} 
\end{subfigure}\hfil
\begin{subfigure}{0.3\textwidth}
\includegraphics[width=0.9\linewidth, height=4cm]{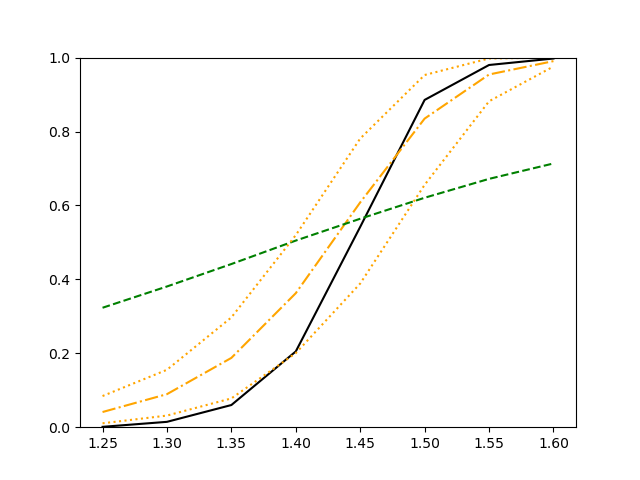}
\end{subfigure}\hfil
\begin{subfigure}{0.3\textwidth}
\includegraphics[width=0.9\linewidth, height=4cm]{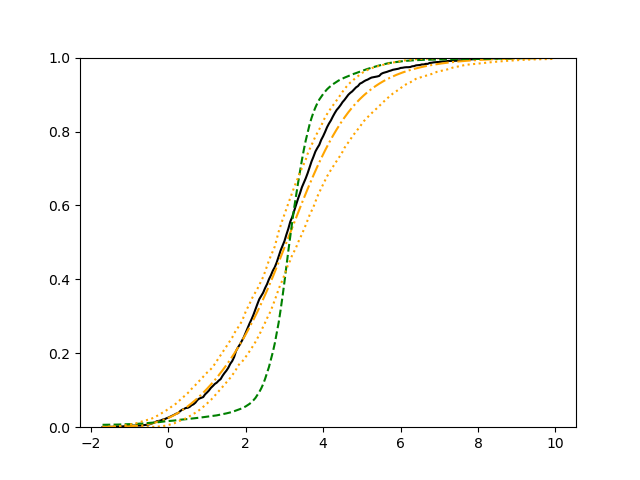}
\end{subfigure}\hfil
\begin{subfigure}{0.3\textwidth}
\includegraphics[width=0.9\linewidth, height=4cm]{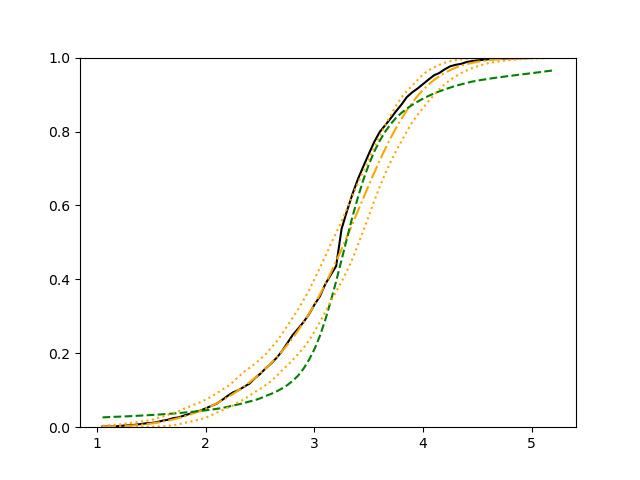}
\end{subfigure}\hfil
\begin{subfigure}{0.3\textwidth}
\includegraphics[width=0.9\linewidth, height=4cm]{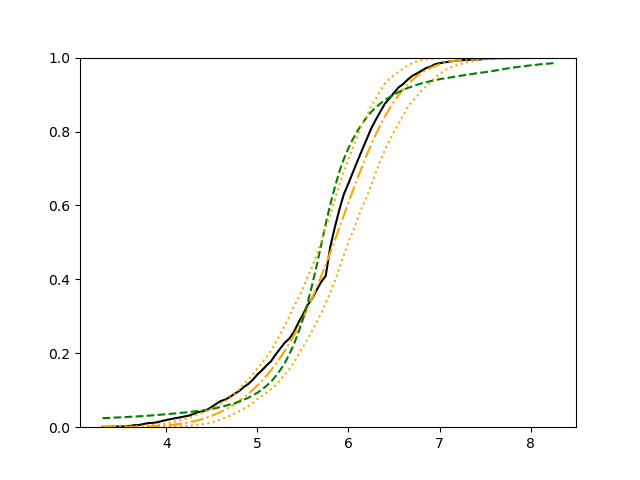} 
\end{subfigure}\hfil
\begin{subfigure}{0.3\textwidth}
\includegraphics[width=0.9\linewidth, height=4cm]{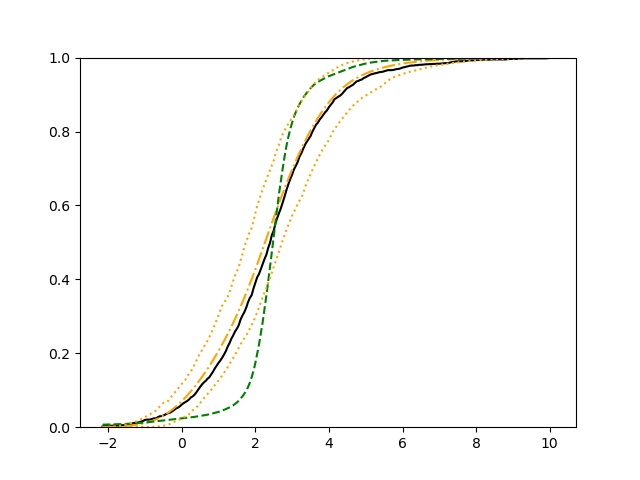} 
\end{subfigure}\hfil
\begin{subfigure}{0.3\textwidth}
\includegraphics[width=0.9\linewidth, height=4cm]{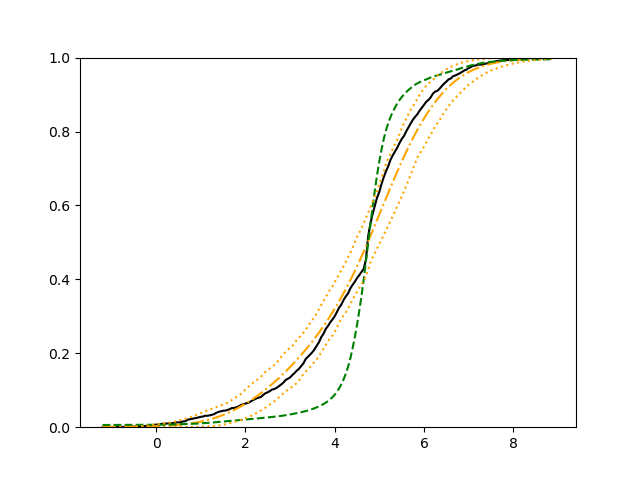} 
\end{subfigure}\hfil
\begin{subfigure}{0.3\textwidth}
\includegraphics[width=0.9\linewidth, height=4cm]{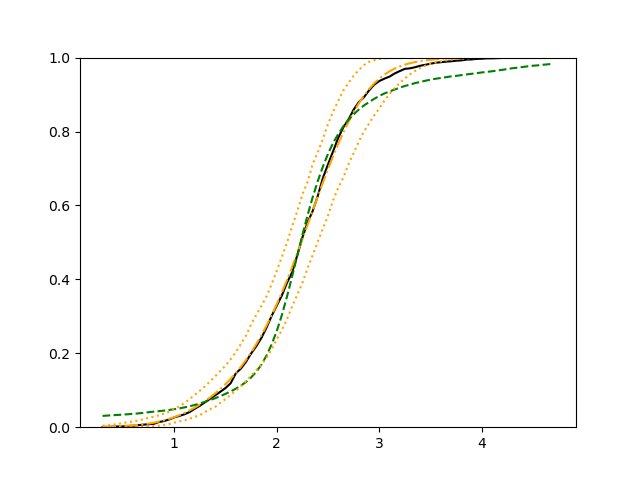} 
\end{subfigure}\hfil
\begin{subfigure}{0.3\textwidth}
\includegraphics[width=0.9\linewidth, height=4cm]{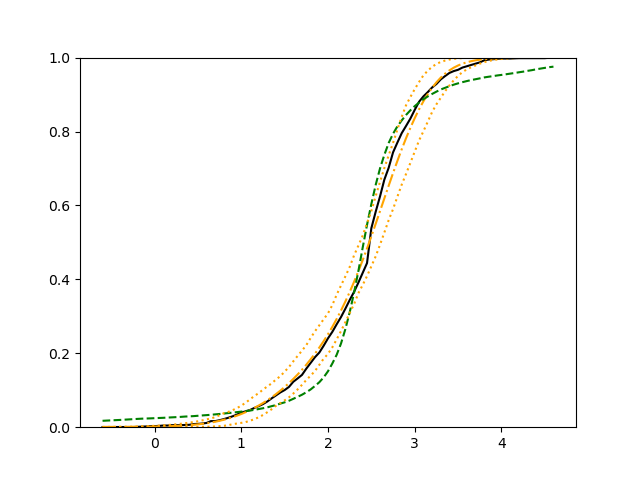} 
\end{subfigure}\hfil
\caption{Conditional distribution functions given 9 different sets of covariate values for uncensored data generated in Setup 2 with c=0.5.}
\label{fig:non_l2}
\end{figure}

For each setup, we generate independent training and validation data sets with equal sample size, then fit both models with our general loss given in (\ref{eq:loss-general}) and the $L_2$ loss using the same neural networks architecture.  Figure~\ref{fig:l2} and Figure~\ref{fig:non_l2} illustrate the comparisons of estimated conditional distribution functions given 9 different sets of covariates between these two methods with a sample size of 5000 and 100 replications.  In these figures, the black solid curve represents the true conditional distribution function, the green dashed curve represents the estimated conditional distribution function using $L_2$ loss, and the orange dash-dot curve represents the estimated conditional distribution using our method. The orange dotted curves are the 90\% confidence band estimated using our method from the 100 repeated experiments.
Figure~\ref{fig:l2} illustrates that  when the error is uncorrelated with the covariates and has constant variance (Setup 1), both methods perform well in estimating the conditional distribution functions. When the error becomes correlated with the covariates and has non-constant variance (Setup 2), the traditional neural networks  method using $L_2$ loss fails, which is illustrated in   Figure~\ref{fig:non_l2}.

We also compare the conditional mean estimates of both methods under two different sample sizes ($n=1000$ and $n=5000$) and two different magnitudes of noises ($c=0.5$ and $c=1$). We evaluate the performance of both methods by averaging the mean and median squared prediction errors, respectively, of 500 independently generated test data points over 100 replications, and summarize the results in Table~\ref{tab:sim general}. Coverage rates of 90\% and 95\% predictive intervals obtained using our method are also presented in Table~\ref{tab:sim general}. In Setup 1, both methods have similar mean squared prediction error, and our method yields slightly smaller median squared prediction error. In Setup 2, our model yields slightly better mean squared error, and significantly better median squared error. Our model provides reasonable prediction coverage rates in both setups,  with improved performance as the sample size increases.

\begin{table}[p]
    \centering
    \begin{tabular}{l c c c c }
    Sample size $n$ & \multicolumn{2}{c}{1000} & \multicolumn{2}{c}{5000} \\ 
    \hline
    Setup 1\\
    $c$ & 0.5 & 1 & 0.5 & 1\\
    \hline
    $L_2$ method mean squared error & 0.50 & 1.83 & 0.45 & 1.73\\
    new method mean squared error & 0.51 & 1.84 & 0.44 & 1.73\\
    $L_2$ method median squared error & 0.17 & 0.59 & 0.14 & 0.52\\
    new method median squared error & 0.16 & 0.56 & 0.13 & 0.51 \\
    new method 90\% coverage rate & 0.90 & 0.90 & 0.90 & 0.90\\
    new method 95\% coverage rate & 0.95 & 0.95 & 0.95 & 0.96\\
    \hline
    Setup 2 \\
    $c$ & 0.5 & 1 & 0.5 & 1\\
    \hline
    $L_2$ method mean squared error & 1.79 & 6.90 & 1.65 & 6.47\\
    new method mean squared error & 1.74 & 6.84 & 1.58 & 6.34\\
    $L_2$ method median squared error & 0.09 & 0.27 & 0.04 & 0.13\\
    new method median squared error & 0.02 & 0.08 & 0.01 & 0.05\\
    new method 90\% coverage rate &  0.87 & 0.87 & 0.91 & 0.91\\
    new method 95\% coverage rate & 0.91 & 0.91 & 0.95 & 0.94\\
    \end{tabular}
    \caption{Average mean/median squared errors of both methods and the prediction coverage rate of the new method over 100 replications.}
    \label{tab:sim general}
\end{table}

\section{Real-World Data Sets}
\label{sec:real data}

\subsection{Censored Data Examples}

There are five real-world data sets analyzed by  \cite{kvamme2019timetoevent}. We re-analyze all these data sets using our method  and compare with  \cite{kvamme2019timetoevent}, except one data set that contains too many ties for which a discrete survival model would be more appropriate.  Theses four data sets are: the Study to Understand Prognoses Preferences Outcomes and Risks of Treatment (SUPPORT), the Molecular Taxonomy of Breast Cancer International Consortium (METABRIC), the Rotterdam tumor bank and German Breast Cancer Study Group (Rot.\& GBSG), and the Assay Of Serum Free Light Chain (FLCHAIN).

The first three data sets are introduced and preprocessed by \cite{deepsurv}. The fourth data set is from the survival package of R (\cite{survival-package}) and preprocessed by \cite{kvamme2019timetoevent}. These four data sets have sample sizes of a few thousand and the covariate numbers range from 7 to 14. The covariates in these data sets are all time-independent.

To compare with their method, we use the same 5-fold cross-validated evaluation criteria described in \cite{kvamme2019timetoevent}, including concordance index (C-index), integrated Brier score (IBS) and integrated binomial log-likelihood (IBLL). The time-dependent C-index \citep{Antolini2005ATD} estimates the probability that the predicted survival times of two comparable individuals have the same ordering as their true survival times,
\begin{equation*}
    \mbox{C-index} = P\{\widehat{S}(T_i | x_i)<\widehat{S}(T_i | x_j) | T_i<T_j,\Delta_i=1\}.
\end{equation*}
The generalized Brier score \citep{Graf1999AssessmentAC} can be interpreted as the mean squared error of the probability estimates. To account for censoring, the scores are weighted by inverse censoring survival probability. In particular, for a fixed time $t$,
\begin{equation*}
    BS(t) = \frac{1}{n}\sum_{i=1}^n \left\{\frac{\widehat{S}(t | x_i)^2I(Y_i \le t, \Delta_i=1)}{\widehat{G}(Y_i)}+ \frac{\left[1-\widehat{S}(t | x_i)\right]^2I(Y_i > t)}{\widehat{G}(t)}
    \right\}.
\end{equation*}
where $\widehat{G}(t)$ is the Kaplan-Meier estimate of the censoring time survival function. The binomial log-likelihood is similar to the Brier score, 
\begin{equation*}
    BLL(t) = \frac{1}{n}\sum_{i=1}^n\left\{\frac{\log\left[1-\widehat{S}(t | x_i)\right]I(Y_i \le t, \Delta_i=1)}{\widehat{G}(Y_i)}+ \frac{\log\left[\widehat{S}(t | x_i)\right]I(Y_i > t)}{\widehat{G}(t)}\right\}.
\end{equation*}
The integrated Brier score IBS and the integrated binomial log-likelihood score IBLL are calculated by numerical integration over the time duration of the test set.

The results of our method are summarized in Table~\ref{tab:survival_real_tuning}, together with the results of \cite{kvamme2019timetoevent} for a comparison. For SUPPORT and METABRIC data, our model yields the best integrated brier score and integrated binomial log-likelihood.

For Rot.\&GBSG data, our model has the best C-index. The other results are comparable to that from the \cite{kvamme2019timetoevent}. Note that in the 5-fold cross validation procedure, we use the set-aside data only as the test set for evaluation of the criteria and the rest of the data for training and validation of the neural networks, whereas \cite{kvamme2019timetoevent} use the set-aside data as both the test set and the validation set which would lead to more favorable evaluations.

\begin{table}[p]
    \centering
    \begin{tabular}{llllllllll}
    & \multicolumn{3}{c}{C-Index} & \multicolumn{3}{c}{IBS} & \multicolumn{3}{c}{IBLL}\\
    & a & b & c & a & b & c & a & b & c\\
    \hline
    SUPPORT & 0.613 & 0.629 & 0.609 & 0.213 & 0.212 & 0.195$^{**}$ & -0.615 & -0.613 & -0.574$^{**}$ \\
    METABRIC & 0.643 & 0.662 & $0.652^*$ & 0.174 & 0.172 & 0.166$^{**}$ & -0.515 & -0.511 & -0.496$^{**}$\\
    Rot.\&GBSG & 0.669 & 0.677 & 0.680$^{**}$ & 0.171 & 0.169 & 0.176 & -0.509 & -0.502 & -0.524\\
    FLCHAIN & 0.793 & 0.790 & 0.788 & 0.093 & 0.102 & 0.102$^*$ & -0.314 & -0.432 & -0.336$^*$                          \\
    \end{tabular}
    \caption{Comparisons of different methods (a: Cox-MLP (CC); b: Cox-Time; c: our new method) for analyzing four real data sets. The result of our method is marked with ** when it outperforms both Cox-MLP(CC) and Cox-Time, and is marked with * when it outperforms one of the models. }
    \label{tab:survival_real_tuning}
\end{table}

\subsection{Uncensored Data}

We use QSAR Fish Toxicity data set \citep{fish} for an illustration. This data set is collected for developing quantitative regression models to predict acute aquatic toxicity towards the fish Pimephales promelas (fathead minnow) on a set of 908 chemicals. Six molecular descriptors (representing the structure of chemical compounds) are used as the covariates and the concentration that causes death in 50\% of test fish over a test duration of 96 hours, called $LC_{50}$ 96 hours (ranges from 0.053 to 9.612 with a mean of 4.064) was used as model response. The 908 data points are curated and filtered from experimental data. The six molecular descriptors come from a variable selection procedure through genetic algorithms. In their original research article, the authors used a $k$-nearest-neighbours (kNN) algorithm to estimate the mean. The data set can be obtained from the machine learning repository of the University of California, Irvine (\url{https://archive.ics.uci.edu/ml/datasets/QSAR+fish+toxicity}).

\begin{figure}[p]
\centering
\begin{subfigure}{0.5\textwidth}
\includegraphics[width=0.8\linewidth, height=5cm]{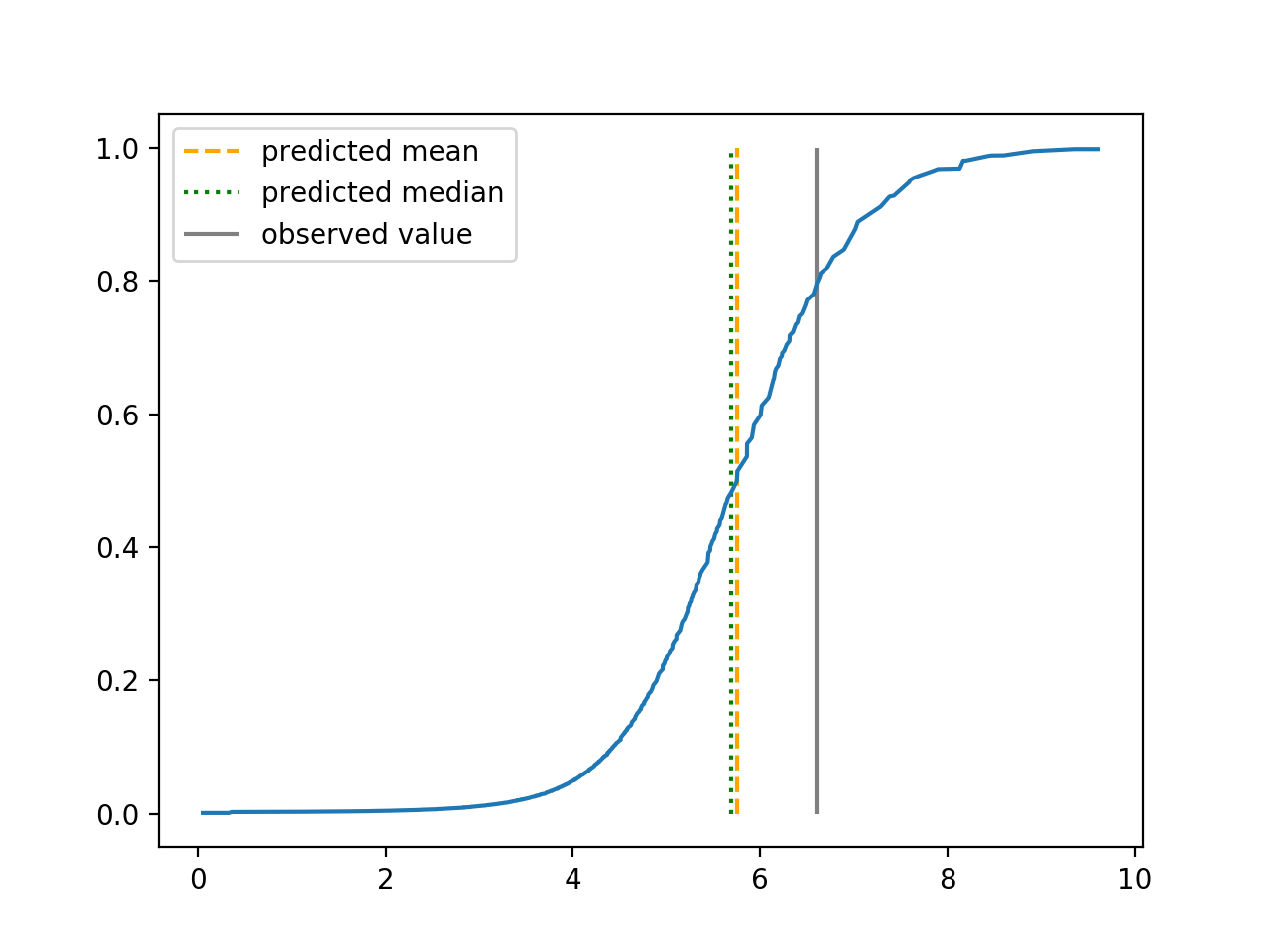} 
\end{subfigure}\hfil
\begin{subfigure}{0.5\textwidth}
\includegraphics[width=0.8\linewidth, height=5cm]{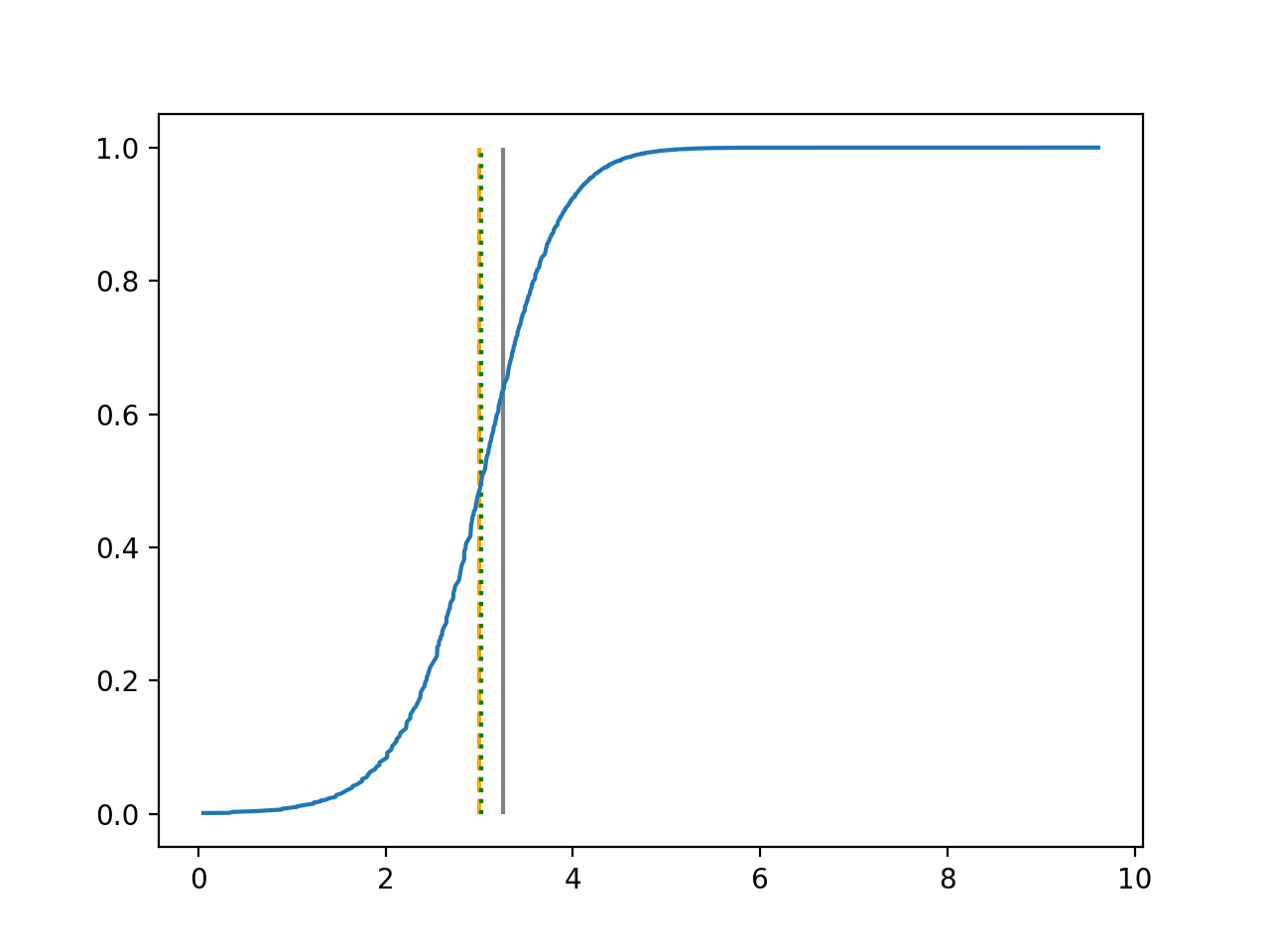}
\end{subfigure}\hfil
\begin{subfigure}{0.5\textwidth}
\includegraphics[width=0.8\linewidth, height=5cm]{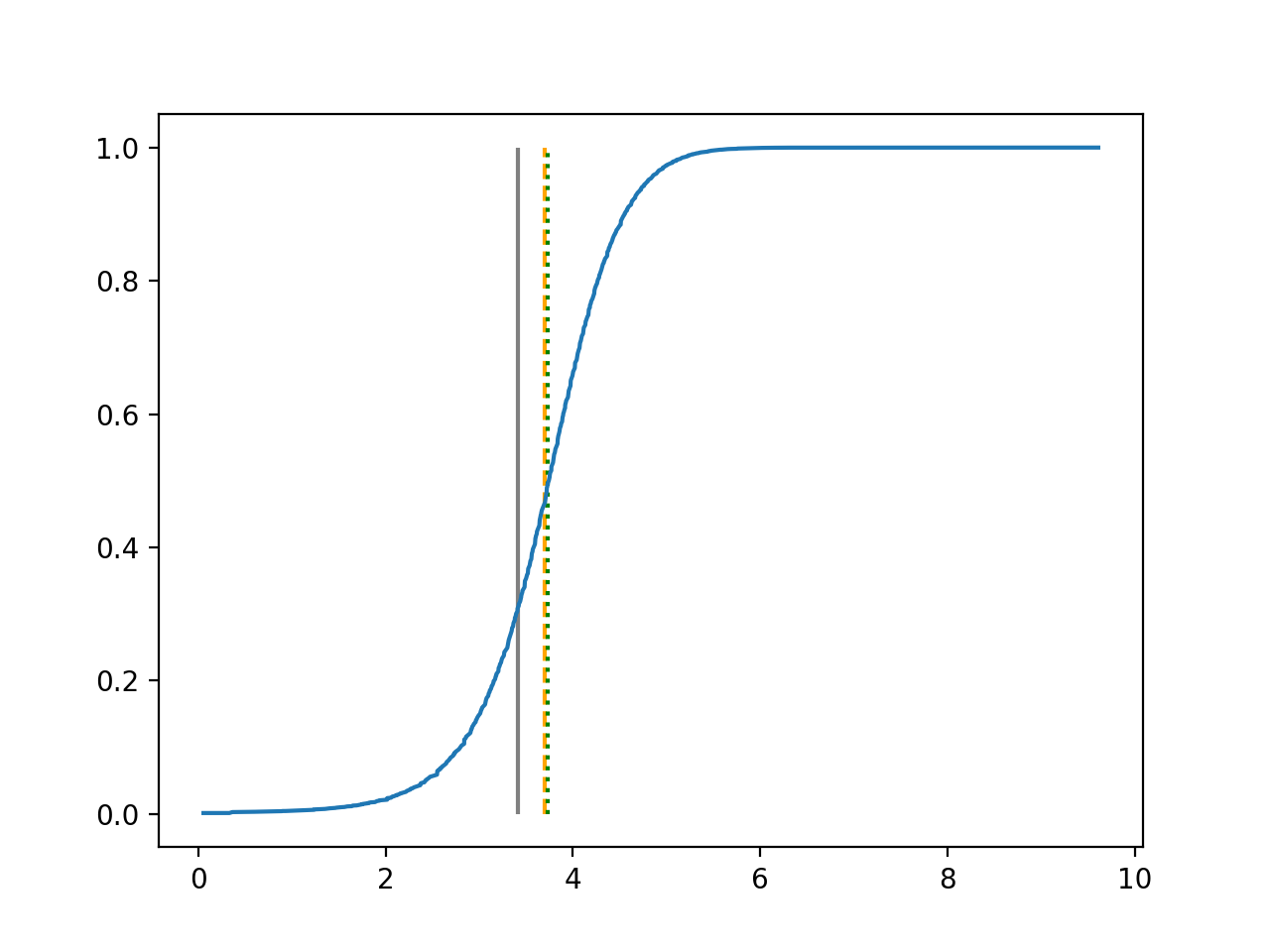}
\end{subfigure}\hfil
\begin{subfigure}{0.5\textwidth}
\includegraphics[width=0.8\linewidth, height=5cm]{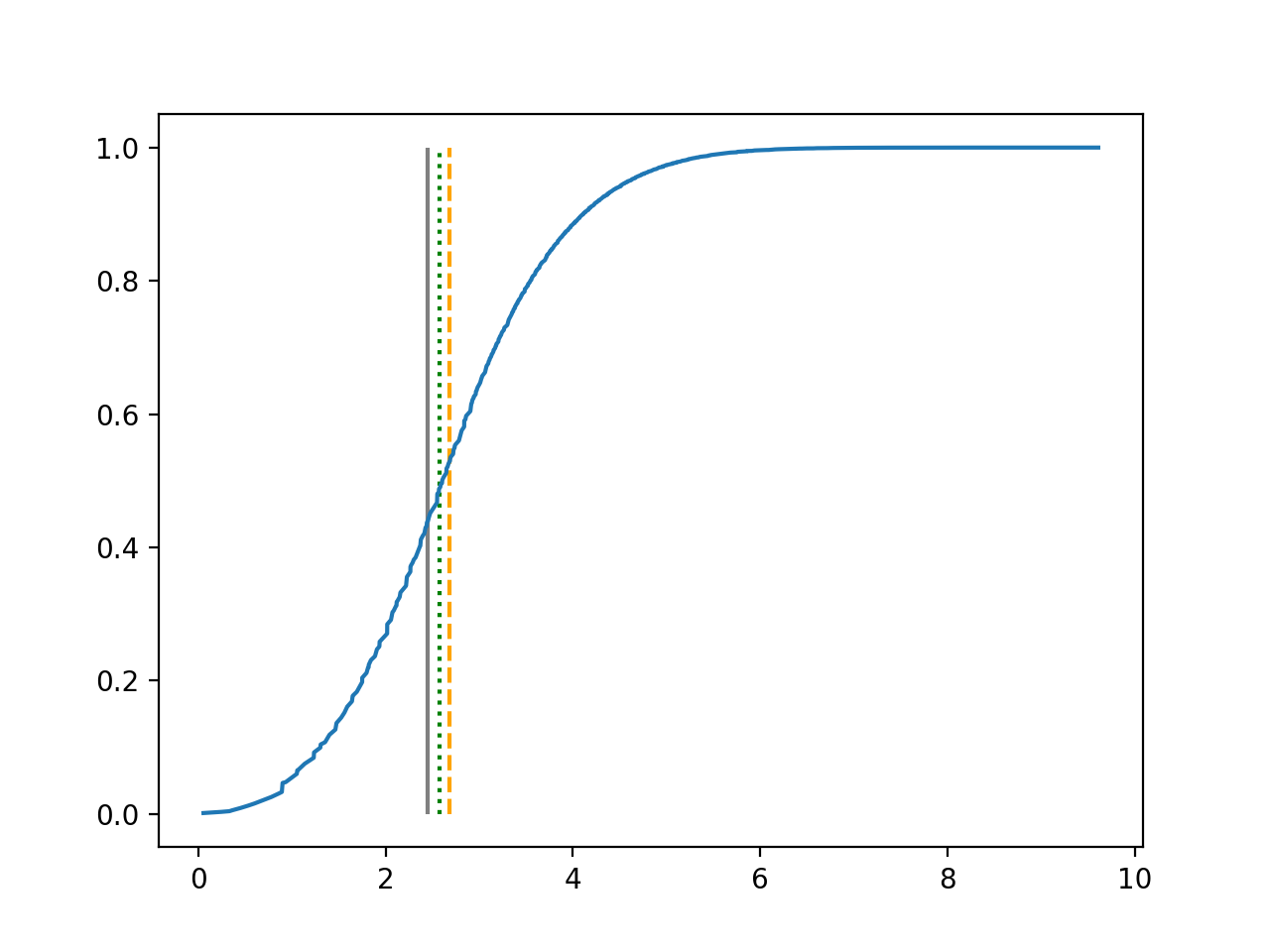}
\end{subfigure}\hfil
\caption{Estimated conditional distribution functions for 4 individuals. The three vertical lines illustrate locations of the predicted mean, the predicted median and the observed value.}
\label{fig:fish_cdf}
\end{figure}

We use 5-fold cross-validated $R^2$, mean squared error and median squared error to evaluate our method and the neural networks with $L_2$ loss on this real-world data set and summarize the results in Table~\ref{tab:general_real}. Our new method yields better prediction in all three criteria. Note that we obtain the same 5-fold cross-validated $R^2$ value of 0.61 as \cite{fish} did by  using kNN. Predicted conditional distribution functions given four different sets of covariate values provided by our method are presented in Figure~\ref{fig:fish_cdf} for an illustration.

\begin{table}[p]
    \centering
    \begin{tabular}{lcc}
         & $L_2$ method & New method \\
         \hline
         Mean squared error & 0.86 & 0.82\\
         Median squared error & 0.22 & 0.20\\
         $R^2$ & 0.59 & 0.61
    \end{tabular}
    \caption{Prediction results of the $L_2$ method and the new method}
    \label{tab:general_real}
\end{table}

\section{Discussion}
\label{sec:discussion}

Early stopping based on a hold-out validation set is used to prevent over-fitting in this work. It is well accepted in deep learning field that early stopping is an eﬀective regularization method. 

\cite{Goodfellow-et-al-2016} pointed out that, in the case of a linear regression model with a quadratic error function and using simple gradient descent, early stopping is equivalent to $L_2$ regularization. 

Thus, intuitively, the validation loss can be a good approximation of the population loss, which implies the estimator obtained using early stopping can be a good approximation of the minimizer of the population loss. A thorough investigation of the asymptotic behavior of our approach using early stopping would be of great interest.

The data expansion technique used in this article provides a simple and natural way of numerically evaluating the full likelihood based loss function. However, it seems that the data expansion would increase the effective sample size from $n$ to $n^2$. This may not be a concern for the survival problem with time-varying covariates because each covariate process needs to be observed at least at all distinct time points, leading to  an order of $n^2$ number of distinct data points. When covariates are random variables other than stochastic processes, the sample size is indeed $n$, thus there should be a large room for developing more efficient numerical approaches. In particular, a recent work by \cite{soden} comes to our attention,  which combines neural networks with an ordinary differential equation (ODE) framework to estimate the conditional survival function given a set of  baseline covariates, in other words, time-independent covariates. They use ODE solver to integrate the cumulative hazard function from an initial value and its derivative (the hazard function). Based on adjoint sensitivity analysis, they are able to avoid going into the ODE solver in back propagation, but use another ODE to calculate the gradient and update the parameters. They show such an algorithm is faster than conventional methods.  It is of great interest to extend the ODE approach to the survival problem with time-varying covariates and the case of arbitrary uncensored data as well to potentially  speed up the computation, which is under current investigation.

\section*{Acknowledgements}

This work was supported in part by NIH R01 AG056764 and NSF DMS 1915711.

\clearpage
\bibliography{Bibliography}

\begin{thebibliography}{24}
\providecommand{\natexlab}[1]{#1}
\providecommand{\url}[1]{\texttt{#1}}
\expandafter\ifx\csname urlstyle\endcsname\relax
  \providecommand{\doi}[1]{doi: #1}\else
  \providecommand{\doi}{doi: \begingroup \urlstyle{rm}\Url}\fi

\bibitem[Antolini et~al.(2005)Antolini, Boracchi, and
  Biganzoli]{Antolini2005ATD}
Laura Antolini, Patrizia Boracchi, and Elia~M. Biganzoli.
\newblock A time-dependent discrimination index for survival data.
\newblock \emph{Statistics in medicine}, 24 24:\penalty0 3927--44, 2005.

\bibitem[Biganzoli et~al.(1998)Biganzoli, Boracchi, Mariani, and
  Marubini]{biganzoli}
Elia Biganzoli, Patrizia Boracchi, Luigi Mariani, and Ettore Marubini.
\newblock Feed forward neural networks for the analysis of censored survival
  data: a partial logistic regression approach.
\newblock \emph{Statistics in Medicine}, 17\penalty0 (10):\penalty0 1169--1186,
  1998.

\bibitem[Brown et~al.(1997)Brown, Branford, and Moran]{brown}
Stephen~F. Brown, Alan~J. Branford, and William Moran.
\newblock On the use of artificial neural networks for the analysis of survival
  data.
\newblock \emph{IEEE Transactions on Neural Networks}, 8\penalty0 (5):\penalty0
  1071--1077, 1997.

\bibitem[Cassotti et~al.(2015)Cassotti, Ballabio, Todeschini, and
  Consonni]{fish}
Matteo Cassotti, Davide Ballabio, Roberto Todeschini, and Viviana Consonni.
\newblock A similarity-based qsar model for predicting acute toxicity towards
  the fathead minnow (pimephales promelas).
\newblock \emph{SAR and QSAR in Environmental Research}, 26\penalty0
  (3):\penalty0 217--243, 2015.

\bibitem[Ching et~al.(2018)Ching, Zhu, and Garmire]{coxnnet}
Travers Ching, Xun Zhu, and Lana~X. Garmire.
\newblock Cox-nnet: An artificial neural network method for prognosis
  prediction of high-throughput omics data.
\newblock \emph{PLOS Computational Biology}, 14\penalty0 (4):\penalty0 1--18,
  04 2018.

\bibitem[Chollet et~al.(2015)]{chollet2015keras}
Francois Chollet et~al.
\newblock Keras, 2015.
\newblock URL \url{https://github.com/fchollet/keras}.

\bibitem[Cox(1972)]{cox}
David~R. Cox.
\newblock Regression models and life-tables.
\newblock \emph{Journal of the Royal Statistical Society: Series B
  (Methodological)}, 34\penalty0 (2):\penalty0 187--202, 1972.

\bibitem[Faraggi and Simon(1995)]{faraggi}
David Faraggi and Richard Simon.
\newblock A neural network model for survival data.
\newblock \emph{Statistics in Medicine}, 14\penalty0 (1):\penalty0 73--82,
  1995.

\bibitem[Gensheimer and Narasimhan(2019)]{nnet}
Michael~F. Gensheimer and Balasubramanian Narasimhan.
\newblock A scalable discrete-time survival model for neural networks.
\newblock \emph{PeerJ}, 7:\penalty0 e6257--e6257, 01 2019.

\bibitem[Giunchiglia et~al.(2018)Giunchiglia, Nemchenko, and van~der
  Schaar]{rnnsurv}
Eleonora Giunchiglia, Anton Nemchenko, and Mihaela van~der Schaar.
\newblock Rnn-surv: A deep recurrent model for survival analysis.
\newblock pages 23--32, 2018.

\bibitem[Goodfellow et~al.(2016)Goodfellow, Bengio, and
  Courville]{Goodfellow-et-al-2016}
Ian Goodfellow, Yoshua Bengio, and Aaron Courville.
\newblock \emph{Deep Learning}.
\newblock MIT Press, 2016.
\newblock \url{http://www.deeplearningbook.org}.

\bibitem[Graf et~al.(1999)Graf, Schmoor, Sauerbrei, and
  Schumacher]{Graf1999AssessmentAC}
Erika Graf, Claudia Schmoor, Willi Sauerbrei, and Martin Schumacher.
\newblock Assessment and comparison of prognostic classification schemes for
  survival data.
\newblock \emph{Statistics in medicine}, 18 17-18:\penalty0 2529--45, 1999.

\bibitem[Hall and Yao(2005)]{hall-aos}
Peter Hall and Qiwei Yao.
\newblock Approximating conditional distribution functions using dimension
  reduction.
\newblock \emph{The Annals of Statistics}, 33(3):\penalty0 1404--1421, 2005.

\bibitem[Hall et~al.(1999)Hall, Wolff, and Yao]{hall-jasa}
Peter Hall, Rodney C.~L. Wolff, and Qiwei Yao.
\newblock Methods for estimating a conditional distribution function.
\newblock \emph{Journal of the American Statistical Association},
  94(445):\penalty0 154--163, 1999.

\bibitem[Harrell et~al.(1984)Harrell, Lee, Califf, Pryor, and Rosati]{c_index}
Frank~E. Harrell, Kerry~L. Lee, Robert~M. Califf, David~B. Pryor, and Robert~A.
  Rosati.
\newblock Regression modelling strategies for improved prognostic prediction.
\newblock \emph{Statistics in medicine}, 3\penalty0 (2):\penalty0 143—152,
  1984.

\bibitem[Kalbfleisch and Prentice(2002)]{Kalbfleisch-Prentice-2002}
John~D. Kalbfleisch and Ross~L. Prentice.
\newblock \emph{The Statistical Analysis of Failure Time Data (2nd ed.)}.
\newblock Hoboken, NJ:Wiley, 2002.

\bibitem[Katzman et~al.(2018)Katzman, Shaham, Cloninger, Bates, Jiang, and
  Kluger]{deepsurv}
Jared~L. Katzman, Uri Shaham, Alexander Cloninger, Jonathan Bates, Tingting
  Jiang, and Yuval Kluger.
\newblock Deepsurv: personalized treatment recommender system using a cox
  proportional hazards deep neural network.
\newblock \emph{BMC Medical Research Methodology}, 18\penalty0 (1):\penalty0
  24, 2018.

\bibitem[Kingma and Ba(2014)]{adam}
Diederik~P. Kingma and Jimmy Ba.
\newblock Adam: A method for stochastic optimization.
\newblock \emph{arXiv preprint}, 2014.
\newblock URL \url{https://arxiv.org/abs/1412.6980}.

\bibitem[Kvamme et~al.(2019)Kvamme, {{\O}}rnulf Borgan, and
  Scheel]{kvamme2019timetoevent}
H{{\aa}}vard Kvamme, {{\O}}rnulf Borgan, and Ida Scheel.
\newblock Time-to-event prediction with neural networks and cox regression.
\newblock \emph{Journal of Machine Learning Research}, 20\penalty0
  (129):\penalty0 1--30, 2019.

\bibitem[Solla et~al.(1988)Solla, Levin, and Fleisher]{Solla1988AcceleratedLI}
Sara~A. Solla, Esther Levin, and Michael Fleisher.
\newblock Accelerated learning in layered neural networks.
\newblock \emph{Complex Syst.}, 2, 1988.

\bibitem[Street(1998)]{street}
Nick~W. Street.
\newblock A neural network model for prognostic prediction.
\newblock In \emph{ICML}, pages 540--546. Citeseer, 1998.

\bibitem[Tang et~al.(2022)Tang, Ma, Mei, and Zhu]{soden}
Weijing Tang, Jiaqi Ma, Qiaozhu Mei, and Ji~Zhu.
\newblock Soden: A scalable continuous-time survival model through ordinary
  differential equation networks.
\newblock \emph{Journal of Machine Learning Research}, 23\penalty0
  (34):\penalty0 1--29, 2022.

\bibitem[Therneau(2021)]{survival-package}
Terry~M. Therneau.
\newblock \emph{A Package for Survival Analysis in R}, 2021.
\newblock URL \url{https://CRAN.R-project.org/package=survival}.
\newblock R package version 3.2-13.

\bibitem[Xiang et~al.(2000)Xiang, Lapuerta, Ryutov, Buckley, and
  Azen]{XIANG2000243}
Anny Xiang, Pablo Lapuerta, Alex Ryutov, Jonathan Buckley, and Stanley Azen.
\newblock Comparison of the performance of neural network methods and cox
  regression for censored survival data.
\newblock \emph{Computational Statistics \& Data Analysis}, 34\penalty0
  (2):\penalty0 243 -- 257, 2000.

\end{thebibliography}

\end{document}